\documentclass[twocolumn]{aastex7}
\usepackage{xcolor}
\usepackage{soul} 
\usepackage{comment}
\definecolor{mypink}{RGB}{255,105,180}

\newcommand\gva{{\sc Genec}}

\newcommand{\Ms}{{\ensuremath{M_{\odot} }}}

\begin{document}

\title{Stellar Yields of Rapidly Rotating Population III Stars for the High Redshift Universe}

\author[orcid=0000-0003-1927-4397,sname=Nandal,gname=Devesh]{Devesh Nandal}
\affiliation{Center for Astrophysics, Harvard and Smithsonian, 60 Garden St, Cambridge, MA 02138, USA}
\email[show]{deveshnandal@yahoo.com}

\author[orcid=0009-0009-0009-1364,sname=Topalakis,gname=Konstantinos]{Konstantinos Topalakis}
\affiliation{Department of Physics, Aristotle University of Thessaloniki, Thessaloniki 54124, Greece}
\email{topalakiskonstantinos@gmail.com}

\author[orcid=0009-0006-2421-5905,sname=Sergienko,gname=Vasilisa]{Vasilisa Sergienko}
\affiliation{Department of Astronomy, University of Virginia, 530 McCormick Rd, Charlottesville, VA 22904, USA}
\email{vasilisa.nikitichna.sergienko@gmail.com}

\author[orcid=0009-0006-2831-2067,sname=Eker,gname=Doga]{Doga Eker}
\affiliation{Department of Astronomy \& Astrophysics, University of Chicago, Chicago, IL 60637, USA}
\email{PUT_EMAIL_HERE}

\author[orcid=0009-0004-5217-4915,sname=Celep,gname=Efe Kurtulus]{Efe K. Celep}
\affiliation{Department of Physics and Astronomy, Carleton College, Northfield, MN 55057, USA}
\email{celepe@carleton.edu}

\author[orcid=0000-0002-4153-053X,sname='Danielle Berg']{Danielle A. Berg}
\affiliation{Department of Astronomy, University of Texas at Austin, 2515 Speedway, Austin, TX 78712, USA}
\affiliation{Cosmic Frontier Center, The University of Texas at Austin, Austin, TX 78712, USA}
\email{PUT_EMAIL_HERE}

\author[orcid=0000-0001-6646-2337,sname=Whalen,gname=Daniel J.]{Daniel J. Whalen}
\affiliation{Institute of Cosmology and Gravitation, Portsmouth University, Dennis Sciama Building, Portsmouth PO1 3FX}
\email{dwhalen1999@gmail.com}

\author[orcid=0000-0003-1927-4397,sname=Tsiatsiou,gname=Sophie]{Sophie Tsiatsiou}
\affiliation{Department of Astronomy, University of Geneva, Chemin Pegasi 51, 1290 Versoix, Switzerland}
\email{sofia.tsiatsiou@unige.ch}

\author[orcid=0000-0001-8764-6522,sname=Hirschi,gname=Raphael]{Raphael Hirschi}
\affiliation{Astrophysics Research Centre, Lennard-Jones Laboratories, Keele University, Keele ST5 5BG, UK}
\affiliation{Kavli IPMU (WPI), University of Tokyo, 5-1-5 Kashiwanoha, Kashiwa 277-8583, Japan}
\email{r.hirschi@keele.ac.uk}

\begin{abstract}
JWST has revealed rapid nitrogen enrichment in high redshift galaxies, renewing the need for stellar yields that follow metal free stars beyond the main sequence and across their final fates. We present Geneva stellar evolution models of rapidly rotating Pop~III stars with $5 \leq M_{\rm ini}/M_\odot \leq 200$ and $\upsilon_{\rm ini}/\upsilon_{\rm crit}=0.7$, evolved to the end of core O-burning. We calculate gross adopted ejecta masses using fate dependent remnant prescriptions spanning core collapse, pulsational pair instability, and complete pair instability. Our quantitative results focus on hydrostatically produced CNO material, while heavier species are treated as final-model reservoirs. Rotation produces primary CNO material by mixing newly synthesized C and O into H-burning layers, where CNO cycling generates nitrogen. The resulting abundance signature depends strongly on which stellar layers escape. Models leaving compact remnants can reach $\log({\rm N/O})=-0.49$ while remaining poor in Si/S/Ar/Ca-rich material. PISN models eject the largest nitrogen masses, $M_{\rm N}=1.8$--$2.5,M_\odot$, but also release $57$--$65,M_\odot$ of oxygen, lowering $\log({\rm N/O})$ to $-1.44$ to $-1.29$. Thus, high nitrogen yields do not imply high N/O ratios. This trend persists after IMF weighting, as increasingly top-heavy populations add oxygen and deep alpha material faster than they build a nitrogen dominated mixture. Our grid provides a new CNO-focused prompt-enrichment dataset for interpreting nitrogen-rich systems including GN-z11, CEERS-1019, and GS~3073.

\end{abstract}

\keywords{\uat{Massive stars}{732} --- \uat{Stellar evolutionary models}{2046} --- \uat{Abundance ratios}{11} --- \uat{Early universe}{435} --- \uat{Chemical abundances}{224} --- \uat{PopIII stars}{1285}}

\section{Introduction}\label{sec:intro}

Population~III (Pop~III) stars mark the first phase of stellar nucleosynthesis in the Universe. They formed from primordial gas and produced the first heavy elements \citep{Bromm2004,Heger2010}. Their remnants also provided early compact objects that could contribute to black-hole (BH) growth and galaxy assembly \citep{Haem2020,Inayoshi2020,latif21a,latif22b}. Pop~III star formation has been studied with cosmological collapse calculations, protostellar accretion models, and radiation-hydrodynamic simulations \citep{Omukai2003,Yoshida2006,Greif2011,Hirano2014}. These studies show that metal-free star formation can produce a broad mass spectrum \citep{Hirano2014,Hirano2015,Susa2019,latif22a}. The final masses depend on accretion history, fragmentation, and radiative feedback \citep{Greif2012,Hosokawa2016,Klessen2023}.  Direct identification of individual Pop~III stars remains difficult, but their chemical imprint can be searched for in metal-poor stars and high-redshift galaxies \citep{jet09b,Heger2010,Kobayashi2023}. Stellar yield calculations therefore remain a central link between first-star theory and observations \citep{Heger2002,Chieffi2004,Limongi2012}.

Primordial stars form by rapid accretion from angular-momentum-rich gas \citep{Greif2012, Stacy2013}, motivating rotation as one of the key physical ingredients in metal-free stellar evolution \citep{Maeder2012,Yoon2012}. Simulations indicate that Pop~III protostars can retain enough angular momentum to rotate at a large fraction of critical velocity \citep{Stacy2011,Stacy2013}. This is because at zero metallicity the metal opacity is absent or strongly reduced, resulting in weak line-driven winds \citep{Kudritzki2002} and inefficient removal of angular momentum \citep{Ekstrom2006,Ekstrom2008}. Rotation can alter the internal structure, enlarge the chemically mixed regions, change the surface composition, and modify the final core masses of massive stars \citep{Maeder2012,Yoon2012,Murphy2021,Tsiatsiou2024}. It can also affect the ionizing output and evolutionary pathway of metal-free stars \citep{Sibony2022}. Even when radiative mass loss is small, rapidly-rotating models can approach critical rotation or the $\Omega\Gamma$ limit \citep{OGlimit2000,Maeder2005,Haemmerle2018,Nandal2025b}. This opens a mechanical mass-loss channel at very low metallicity \citep{Ekstrom2006,Ekstrom2008}.

The nucleosynthetic consequences of rotation are especially important at zero metallicity ($Z=0$). Non-rotating Pop~III yield grids established the basic outcomes of core-collapse and hypernova enrichment \citep{Chieffi2004,Tominaga2007,Limongi2012,chen14a}. Pair-instability (PI) calculations showed that very massive metal-free stars can produce strong even-$Z$ products and distinctive abundance patterns \citep{Heger2002}. Pulsational pair-instability (PPI) models further showed how massive stars can eject part of their envelopes before final collapse \citep{wet13d,chen14a,Woosley2017}. Rotation adds another nucleosynthetic channel. It can mix newly synthesized C and O from He-burning regions into H-burning layers \citep{Ekstrom2008,Murphy2021, Tsiatsiou2024}. CNO cycling in those layers then produces primary N and related CNO isotopes \citep{Chiappini2006,Hirschi2008}. Recent calculations have extended rotating metal-poor and metal-free models toward more advanced burning stages and larger reaction networks \citep{Roberti2024,Griffiths2025}, but a broad yield set for rapidly-rotating Pop~III stars is still needed. This is especially true for models followed to advanced evolutionary stages, with adopted ejecta tabulated over many elements.

This need has become sharper with James Webb Space Telescope (JWST). Several early galaxies show abundance ratios that are difficult to interpret with standard enrichment channels alone. GN-z11 shows strong nitrogen emission and a high inferred N/O ratio at $z\simeq 10.6$ \citep{Bunker2023,Cameron2023}. CEERS-1019 at redshift $z=8.68$ is another extreme N-emitter with measured C, N, O, and Ne abundances \citep{Marques2023}. GS~3073 appears to be even more N-dominated relative to O \citep{Ji2024,Ji2026}. GN-z9p4 and RXCJ2248-ID extend the sample of compact high-redshift N-emitting galaxies \citep{Schaerer2024,Topping2024, Berg2026}. Other JWST abundance measurements now provide C/O, Ne/O, and O/H constraints at comparable epochs \citep{Arellano2022,Arellano2025,Hsiao2025}. These systems do not point to a single enrichment source. They instead require yield tables that can separate primary N production from the release of oxygen-rich and deeper alpha-rich material. Rapidly-rotating Pop~III stars have been proposed as contributors to nitrogen-rich high-redshift systems \citep{Nandal2024f}. Very massive and supermassive Pop~III stars provide another possible channel for the most extreme abundance ratios \citep{Denissenkov2014,Nandal2025b,Nandal2025}.

In this paper, we present a grid of rapidly-rotating Pop~III star models with $5 \leq M_{\rm ini}/M_\odot \leq 200$ and $\upsilon_{\rm ini}/\upsilon_{\rm crit}=0.7$. The models are evolved with the Geneva stellar evolution code ({\gva}) from the zero-age main sequence (ZAMS) through core O-burning where applicable, while the lowest-mass models are followed to their latest stable evolutionary endpoints. We convert the final structures of stars into ejecta yields by first classifying the final fate of each model as core collapse, PPISN, or PISN, and then applying a remnant-mass prescription appropriate to that regime. The resulting tables give gross adopted ejecta masses, with the physical interpretation centered on CNO and with explosion-sensitive heavier species explicitly identified as stage-dependent reservoirs. Section~\ref{sec:2} describes the stellar models, reaction network, and yield prescription. Section~\ref{sec:results} presents the evolution, final structures, mass budget, and individual elemental yields. Section~\ref{sec:Discussion} compares the resulting abundance ratios to high-redshift constraints, explores IMF-weighted mixtures, and discusses the sensitivity to the remnant prescription. Section~\ref{sec:Conclusions} summarizes the main implications for early chemical enrichment.

\section{Methods}\label{sec:2}

The stellar models were computed with {\gva}. We considered single Pop~III star models with initial masses in the range $5 \leq M_{\rm ini}/M_\odot \leq 200$. The models were initialized with a primordial composition ($Z=0$) and evolved from the ZAMS until the end of core O-burning, defined by the central $^{16}O$ abundance, as $X_{\rm c}(^{16}{\rm O}) < 10^{-3}$. The input physics follows {\gva} framework used for massive, rotating, and zero-metallicity stellar models, including the treatment of mixing and angular-momentum transport described in previous {\gva} studies \citep{Eggenberger2008,Nandal2024b, Nandal2025b}. In this work, the models do not include the effects of internal magnetic fields.

The treatment of rotation follows the shellular-rotation formalism used in {\gva}. During core H-burning, which accounts for approximately 90\% of the stellar lifetime, angular momentum is transported through the combined action of meridional circulation and shear according to an advective--diffusive equation,
\begin{equation}
\rho\frac{{\rm d}}{{\rm d}t}\left(r^2\Omega\right)_{M_r}
=
\frac{1}{5r^2}\frac{\partial}{\partial r}
\left(\rho r^4\Omega U_2\right)
+
\frac{1}{r^2}\frac{\partial}{\partial r}
\left(\rho D_{\rm shear}r^4\frac{\partial\Omega}{\partial r}\right),
\end{equation}
where $U_2$ is the radial component of the meridional-circulation velocity. After central H exhaustion, the much shorter evolutionary timescales motivate the diffusive treatment adopted for the advanced burning stages \citep{Ekstrom2008}. The transport of chemical species is treated diffusively throughout, using the combined coefficients $D_{\rm shear}+D_{\rm eff}$ \citep{Ekstrom2012}. This distinction is important because the standard MESA implementation treats both angular momentum and chemical transport in the diffusion approximation \citep{Paxton2013}.

Nuclear burning was followed with the extended {\gva} reaction-network framework used for advanced massive-star evolution. During core H and He burning, the larger light element network follows the reactions required to capture the CNO cycles and the production of primary C, N, and O. From core C burning onward, we use the reduced 25-species GeValNet25 network, which extends the {\gva} $\alpha$-chain backbone and includes an effective electron capture chain from $^{56}{\rm Ni}$ to $^{56}{\rm Cr}$ \citep{Griffiths2025}. GeValNet25 was designed to reproduce the energetics, core structure, and approximate neutronization required for advanced stellar evolution, and its results have been compared directly with MESA models using the analogous reduced \texttt{approx21\_plus\_co56} network \citep{Griffiths2025}. We also employ the updated equation of state implemented by \citet{Griffiths2025}, which couples the original {\gva} equation of state with the Timmes Helmholtz equation of state \citep{Timmes2000}. The present network is therefore appropriate for the stellar evolution and hydrostatic CNO production studied here, but it is not intended to replace an extended network for detailed explosive or Fe-group nucleosynthesis. We consequently interpret the heavier CC and PPISN abundances only as stage dependent final model reservoirs.

At $Z=0$, we do not include radiative driven line winds. Mass loss is therefore limited to the mechanical component that occurs when the surface approaches critical rotation. In such a case, the excess mass is removed so that the surface remains sub-critical. The critical velocity is computed with the $\Omega\Gamma$ prescription of \citet{OGlimit2000}, following the implementation used in previous {\gva} Pop~III rotating models \citep{Georgy2013b,Tsiatsiou2024}.

\subsection{Adopted remnant boundaries and ejecta prescriptions}
\label{sec:yields_method}

We assign fate dependent remnant boundaries and calculate gross ejected masses from the final hydrostatic structures. These adopted remnant masses are not intended as first principles predictions of the compact object masses\citep{Rollinde2009}. Because the models are not followed through collapse and explosion, the ejecta--remnant partition is assigned parametrically. We first define the hydrogen-free and helium-free core masses from the final abundance profiles as the enclosed mass coordinates where $X<10^{-2}$ and $Y<10^{-2}$, respectively. We denote these by $M_{\rm He}$ and $M_{\rm CO}$ and use them to classify each model into the core-collapse, pulsational pair-instability, or pair-instability channel, following the standard core-mass framework of \citet{Heger2002} and \citet{Woosley2017}.

For models assigned to the core-collapse branch, we determine the remnant mass from an energy-limited mass cut. For a shell of mass $\Delta m_j$ at radius $r_j$ enclosing mass $M_{r,j}$, the gravitational binding contribution is approximated as
\begin{equation}
\Delta E_{{\rm grav},j} \simeq \frac{G M_{r,j}\,\Delta m_j}{r_j},
\end{equation}
so that the energy required to unbind all material exterior to a trial mass cut $m$ is
\begin{equation}
E_{\rm bind}(>m)=\sum_{M_{r,j}>m}\Delta E_{{\rm grav},j}.
\end{equation}
The remnant mass is then the deepest mass coordinate for which the adopted explosion energy, $E_{\rm exp}$, can still unbind the overlying layers,
\begin{equation}
M_{\rm rem} \equiv m \ \ {\rm such\ that}\ \ E_{\rm bind}(>m)\leq E_{\rm exp}.
\end{equation}
Our fiducial grid adopts a purely gravitational criterion with $E_{\rm exp}=1~{\rm B}$, where $1~{\rm B}=10^{51}$~erg. We also evaluated internal-energy and enthalpy corrections as systematic tests of the mass cut, restricting these thermal-support terms to cool, weakly degenerate layers; the resulting variants are used only to estimate the uncertainty in $M_{\rm rem}$ and not in the fiducial yield tables. The ejected mass of species $i$ is then
\begin{equation}
M_{i,{\rm ej}}=\sum_{M_{r,j}>M_{\rm rem}} X_{i,j}\,\Delta m_j .
\end{equation}

Models in the pulsational pair-instability regime are not treated with the core-collapse mass-cut prescription. Instead, we assign the final remnant mass by
\begin{equation}
M_{\rm rem}^{\rm PPISN}=\alpha_{\rm P}\,M_{\rm rem,no\,PSN},
\end{equation}
where $\alpha_{\rm P}$ is the pair-instability reduction factor \citep[Equation B2 of][]{Spera2017} and $M_{\rm rem,no\,PSN}$ is the remnant mass in the absence of pulsational mass loss. The ejecta are then estimated as the hydrostatic reservoir above the fitted remnant mass, 
\begin{equation}
M_{i,{\rm ej}}^{\rm PPISN}=\sum_{M_{r,j}>M_{\rm rem}^{\rm PPISN}} X_{i,j}\,\Delta m_j .
\end{equation}
This treatment captures the net mass budget implied by pulsational stripping but does not follow the pulse sequence or pulse-by-pulse nucleosynthesis.

For models in the full pair-instability regime, we set
\begin{equation}
M_{\rm rem}^{\rm PISN}=0
\end{equation}
and compute the ejecta as the sum of an explosive He core contribution and the hydrostatic envelope exterior to that core. The explosive component is obtained by interpolating the He core yields of \citet{Heger2002} in $M_{\rm He}$,
\begin{equation}
M_{i,{\rm core}}^{\rm PISN}={\cal I}_i(M_{\rm He}),
\end{equation}
where ${\cal I}_i$ denotes the interpolation of their tabulated stable post-decay yields. Because those calculations are given for exploding He cores, we add the composition of the overlying envelope in the full stellar model,
\begin{equation}
M_{i,{\rm env}}^{\rm PISN}=\sum_{M_{r,j}>M_{\rm He}} X_{i,j}\,\Delta m_j ,
\end{equation}
and take the total ejecta to be
\begin{equation}
M_{i,{\rm ej}}^{\rm PISN}=M_{i,{\rm core}}^{\rm PISN}+M_{i,{\rm env}}^{\rm PISN}.
\end{equation}

\begin{table*}
\centering
\caption{Summary of the fate-dependent ejecta prescriptions. The He and CO core masses classify the final fate but are not used to interpolate the CC or PPISN elemental ejecta.}
\label{tab:ejecta_methods}
\resizebox{\textwidth}{!}{
\begin{tabular}{lllll}
\hline
Regime & Fate classification & Adopted remnant boundary & Composition used for ejecta & External yield interpolation \\
\hline
CC &
Final core structure &
$E_{\rm bind}(>m)\leq 1\,{\rm B}$ &
GENEC profile above $M_{\rm rem}$ plus winds &
None \\

PPISN &
He-core mass &
Spera et al. prescription &
GENEC profile above $M_{\rm rem}^{\rm PPISN}$ plus winds &
None \\

PISN &
He-core mass &
$M_{\rm rem}=0$ &
Heger--Woosley explosive He core plus GENEC envelope &
PISN He core only \\
\hline
\end{tabular}}
\end{table*}

For the CC models, we assume that material exterior to the adopted binding-energy boundary is ejected without further nuclear processing. These ejecta therefore do not include shock-induced explosive nucleosynthesis, multidimensional mixing, or subsequent fallback \citep{jet09b}. The CNO results are the most robust quantities for the present purpose because these elements are produced predominantly during hydrostatic H- and He-burning and reside largely outside the deepest shock-processed layers. Previous GENEC calculations found that elements lighter than Ne are modified only marginally by explosive nucleosynthesis, while a comparison between the end of core O- and core Si-burning found changes below approximately 15\% for C, N, O, and Ne in the relevant benchmark model \citep{Hirschi2008,Ekstrom2008}. Species heavier than Ne in the CC and PPISN models are retained in the tables as stage-dependent final-model reservoirs and should not be interpreted as final explosion yields. For the PISN models, the explosive He-core contribution is taken from the detailed post-decay calculations of \citet{Heger2002}, while the composition of the rotating overlying envelope is calculated with GENEC.

A sometimes misunderstood point in the stellar evolution community is that the vast majority of 30 - 90 \Ms\ Pop III stars collapse to BHs without explosion because the sheer mass of the outer layers of the star overwhelm the central engines of CC SNe -- core collapse and bounce.  Hypernovae and gamma-ray bursts \citep[GRBs;][]{nom06,nom10,smidt13a,mes13a} are exceptions to this rule but only comprise $\sim$ 1 in 10,000 of all core-collapse events and are not generally thought to contribute significantly to primordial cosmological enrichment \citep[e.g.,][]{jet09b}.  Nevertheless, for completeness we calculate yields with and without explosions in this mass range, tallying yields produced only by mass loss prior to collapse in the latter case.

The wider GENEC framework has been tested against both early- and late-stage observables, including main-sequence widths, surface abundances, rotational velocities, red-supergiant locations, and supernova progenitor properties, while advanced very-low-metallicity models reproduce observed CNO abundance trends in extremely metal-poor stars \citep{Georgy2013a,Hirschi2008}.


\begin{figure*}[t]
    \centering
    \includegraphics[width=0.49\textwidth]{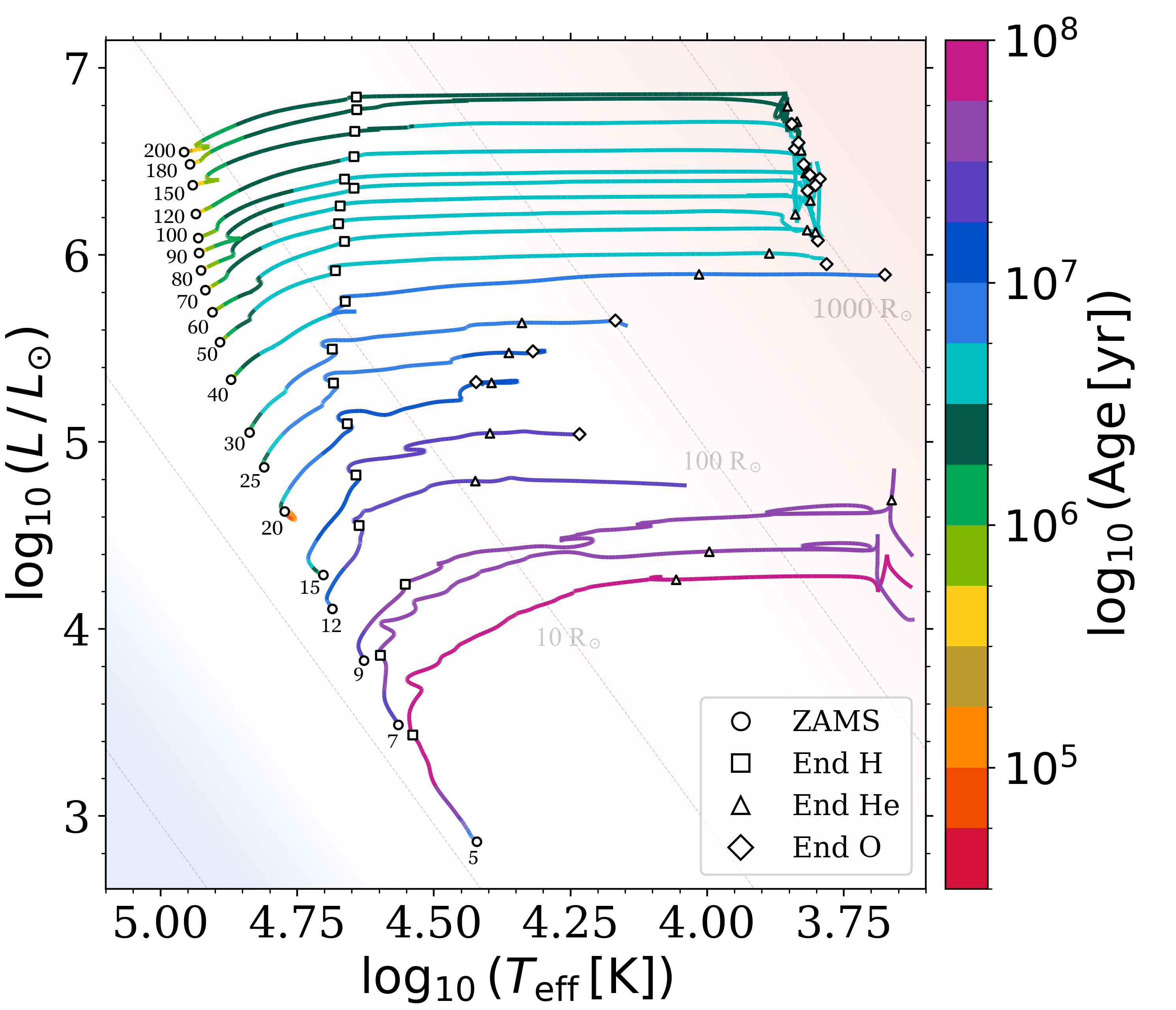}
    \hfill
    \includegraphics[width=0.49\textwidth]{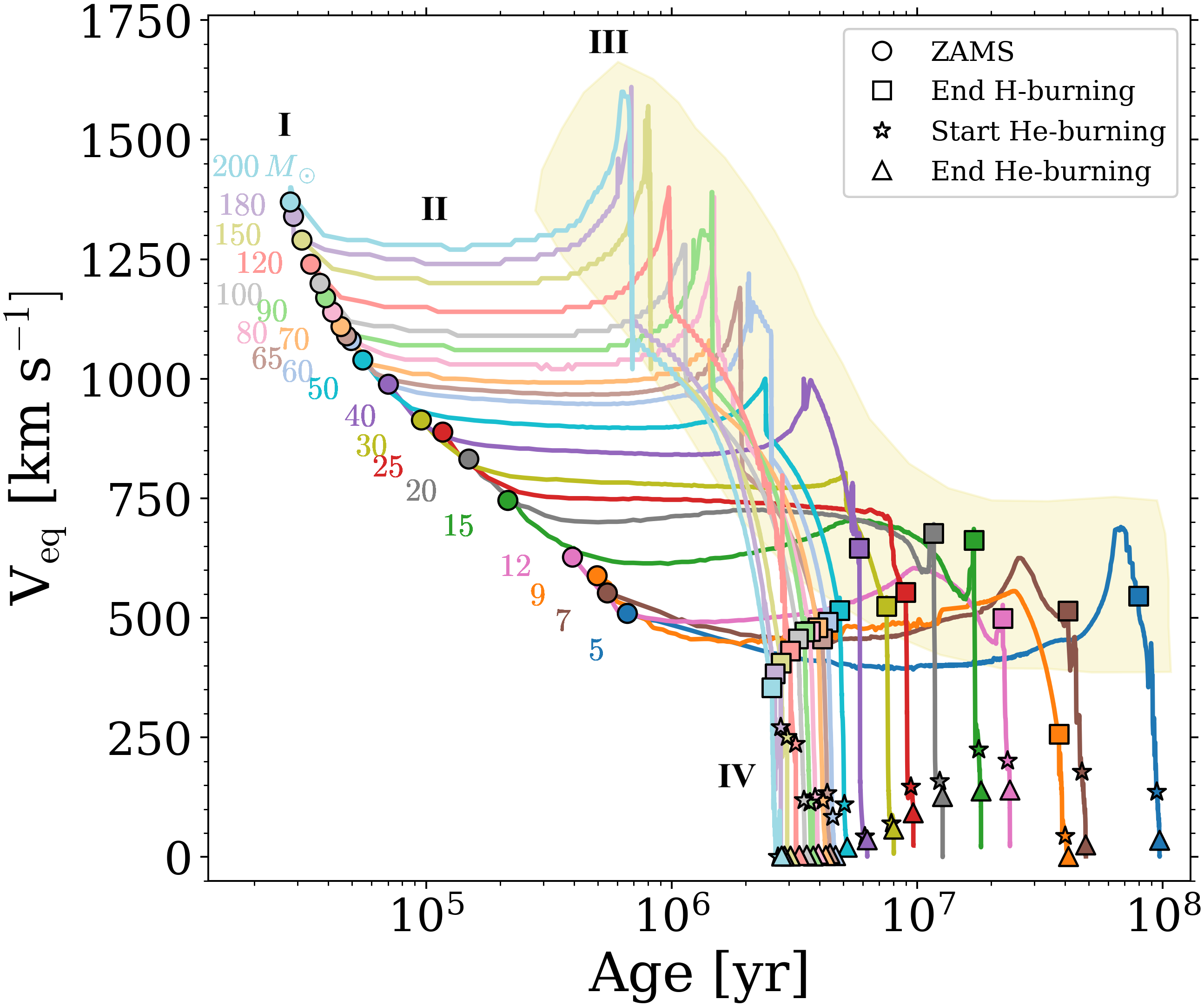}
    \caption{Left: Hertzsprung--Russell diagram of the $Z=0$, $\upsilon_{\rm ini}/\upsilon_{\rm crit}=0.7$ models from $5$ to $200\,M_\odot$, color-coded by stellar age. Circles, squares, triangles, and diamonds mark the ZAMS, end of core-H burning, end of core-He burning, and end of core-O burning, respectively. Right: surface equatorial velocity versus age for the same models. In this panel, circles, squares, diamonds, and triangles denote the ZAMS, end of core-H burning, start of core-He burning, and end of core-He burning, respectively. The shaded region marks the adopted near-critical $\Omega\Gamma$ domain.}
    \label{fig:hrd_veq}
\end{figure*}

\section{Results}\label{sec:results}

\subsection{Global evolutionary behavior of the model grid}
\label{sec:global_evolution}

We summarize the global evolution of the $Z=0$, $\upsilon_{\rm ini}/\upsilon_{\rm crit}=0.7$ grid in Figure~\ref{fig:hrd_veq}. In the left panel, a quasi-chemically homogeneous signature is identified by a blueward drift toward higher $T_{\rm eff}$ during core-H burning. This behavior is strongest at the low-mass end. The $5\,M_\odot$ model moves from $\log(T_{\rm eff}/{\rm K})\simeq 4.42$ on the ZAMS to $\simeq 4.54$ during the MS, and the $7\,M_\odot$ model shows a similar excursion. Between $9$ and $20\,M_\odot$, the same trend is still present, but it lasts for a progressively smaller fraction of core-H burning as the mass increases. Above $20\,M_\odot$, the blueward migration disappears and the tracks evolve redward throughout core-H burning. We therefore identify the $5$ and $7\,M_\odot$ models as the closest cases to quasi-homogeneous evolution, while the $9$--$20\,M_\odot$ models show only a partial and temporary tendency in that direction. This mass dependence likely reflects the shorter mixing path lengths in the more compact low-mass stars. At higher mass, rotational transport can no longer erase the growing chemical stratification. The later development of an extended intermediate convective zone around the H-burning shell then drives the evolution farther from a homogeneous channel \citep{Ekstrom2008,Tsiatsiou2024}.
  
Surface-rotation evolution is shown in the right panel of Figure~\ref{fig:hrd_veq}, with four distinct phases marked. Phase~I is the ZAMS, where the equatorial velocity ($V_{\rm eq}$) rises monotonically with mass from $512~{\rm km\,s^{-1}}$ at $5\,M_\odot$ to $990~{\rm km\,s^{-1}}$ at $40\,M_\odot$ and $1.37\times10^{3}~{\rm km\,s^{-1}}$ at $200\,M_\odot$. Since all models start at the same fraction of the critical velocity, this trend reflects the increase of $\upsilon_{\rm crit}$ with respect to initial mass. Phase~II is the broad decline in $V_{\rm eq}$ seen in all tracks. At $Z=0$, weak core-envelope coupling and very small radiative mass loss keep the surface layers close to local angular-momentum conservation. As the stellar radius grows during the MS phase, the surface therefore spins down \citep{Ekstrom2008,Murphy2021}. Phase~III is the approach to the $\Omega\Gamma$ limit and is marked by the gold shading in Figure~\ref{fig:hrd_veq}. Using $\Omega\Gamma \ge 0.95$, the main near-critical episode extends from $6.12\times10^{7}$ yr to $7.74\times10^{7}$ yr in the $5\,M_\odot$ model, from $3.28\times10^{6}$ yr to $5.82\times10^{6}$ yr in the $40\,M_\odot$ model, and from $5.63\times10^{5}$ yr to $6.84\times10^{5}$ yr in the $200\,M_\odot$ model. These intervals correspond to $16.7\%$, $41.0\%$, and $4.2\%$ of the total computed evolution. The fractional duration is therefore not monotonic with mass and reaches a maximum at intermediate mass in this representative set. After core-H exhaustion, all models undergo a final spin-down driven by envelope expansion denoted by Phase~IV. The subsequent advanced burning stages proceed at very low surface rotation. At the end of core-O burning, the $5$, $40$, and $200\,M_\odot$ models have $V_{\rm eq}=0.84$, $1.11$, and $2.44~{\rm km\,s^{-1}}$, respectively. This mass dependence is one of the ingredients that later produces the non-monotonic behavior in the integrated yields and abundance ratios.

\begin{figure}[t]
    \centering
    \includegraphics[width=\columnwidth]{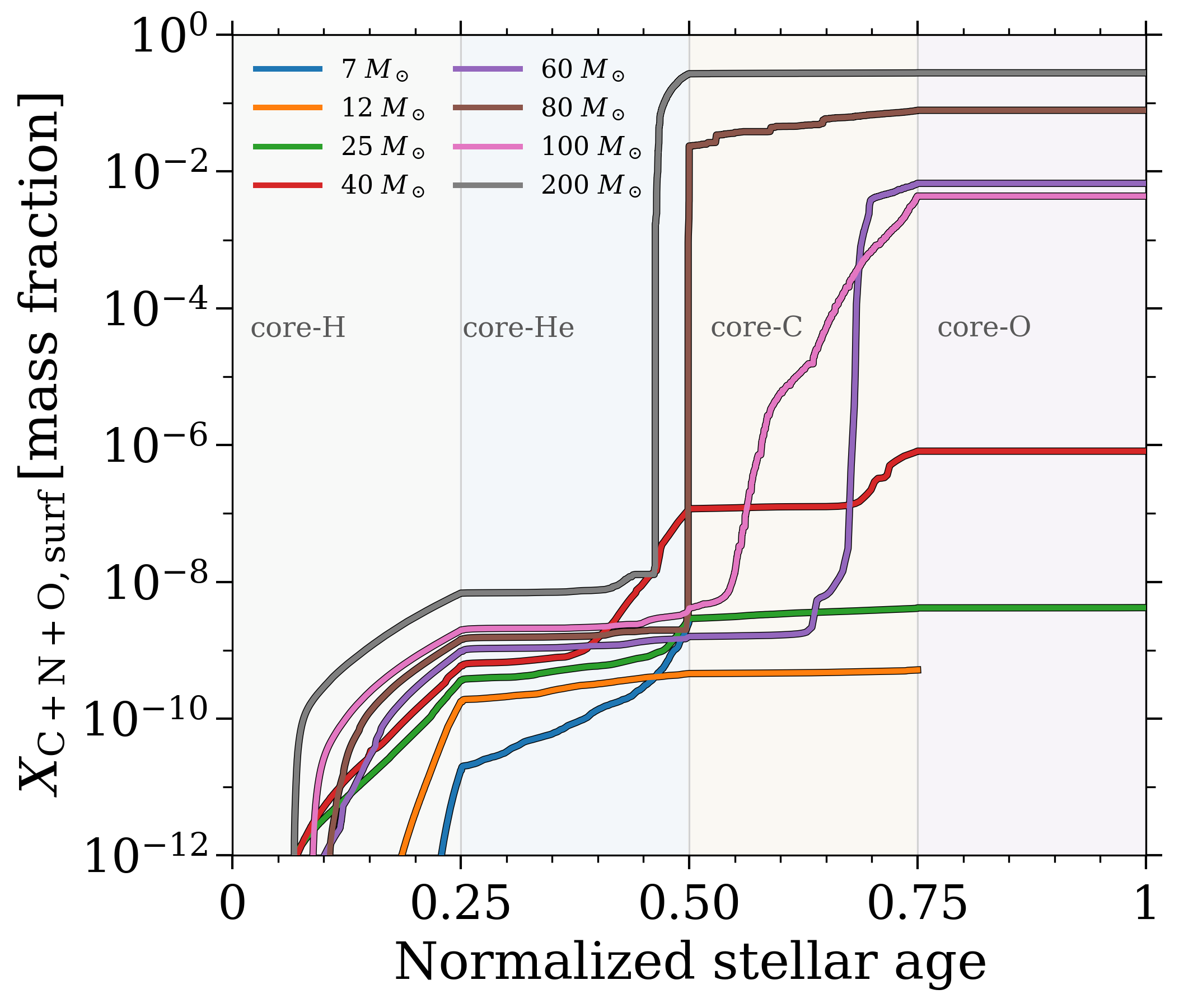}
    \caption{Surface CNO mass fraction as a function of fraction of stellar age for selected models in the $Z=0$, $\upsilon_{\rm ini}/\upsilon_{\rm crit}=0.7$ grid. Only a subset of masses is shown for clarity. The shaded regions mark the phases of core-H, core-He, core-C, and core-O burning, from left to right.}
    \label{fig:surface_cno}
\end{figure}

\subsection{Surface CNO enrichment}
\label{sec:surface_cno}

Figure~\ref{fig:surface_cno} shows the evolution of the total surface CNO mass fraction, defined as the sum of the $^{12}$C, $^{13}$C, $^{14}$N, $^{16}$O, $^{17}$O, and $^{18}$O mass fractions, through the four normalized burning intervals. During core-H burning, all models remain extremely CNO poor at the surface, with $X_{\rm CNO}({\rm surf})\lesssim 10^{-8}$ and most tracks near $10^{-10}$--$10^{-9}$. This is expected for Pop~III stars, which begin without CNO catalysts. Their compact structure also limits core-envelope coupling, so the newly produced nuclei do not reach the surface efficiently.

The main change begins after core-He ignition, when fresh $^{12}$C and $^{16}$O can be transported into the H-burning shell. The surface response is strongly mass dependent. The $200\,M_\odot$ and $80\,M_\odot$ models enrich first and most strongly, reaching $X_{\rm CNO}({\rm surf})\sim2\times10^{-1}$ and $\sim8\times10^{-2}$ near the end of core-He burning. The $100\,M_\odot$ model also enriches strongly, but rises more gradually and saturates near $\sim4\times10^{-3}$. The $60\,M_\odot$ model delays its main jump until late core-C burning and levels off near $\sim7\times10^{-3}$. By contrast, the $40\,M_\odot$ model reaches only $\sim10^{-6}$, while the $25\,M_\odot$ and lower-mass tracks remain below a few $10^{-9}$. Both the onset and the amplitude of the surface enrichment are therefore non-monotonic. Rotation is required to move C- and O-rich material out of the He-burning core, but rapid rotation alone does not guarantee strong surface enrichment. At $Z=0$, meridional circulation is weak and direct homogenization is limited. Large surface jumps appear only when rotational transport also strengthens the H-burning shell and enables deep convective coupling to the overlying envelope. When that response is weaker or delayed, the surface enrichment remains smaller and more diluted \citep{Ekstrom2008,Maeder2012, Hirschi2008,Murphy2021,Tsiatsiou2024}.

After the main rise, many tracks flatten because later burning mainly redistributes material within the CNO group. The plateau values still differ by many orders of magnitude because the enriched material is not mixed through the same envelope mass in every model. Early and deep coupling exposes a larger CNO-rich reservoir, whereas weak or delayed coupling leaves a more pristine H-rich envelope. The lowest-mass models shown do not reach core-O burning because they develop degenerate cores. This mass dependence is one of the ingredients that later produces the non-monotonic behavior seen in the integrated yields and abundance ratios.


\subsection{Final shell structure of the $60\,M_\odot$ model}
\label{sec:onion_60}

\begin{figure*}
    \centering
    \includegraphics[width=0.99\linewidth]{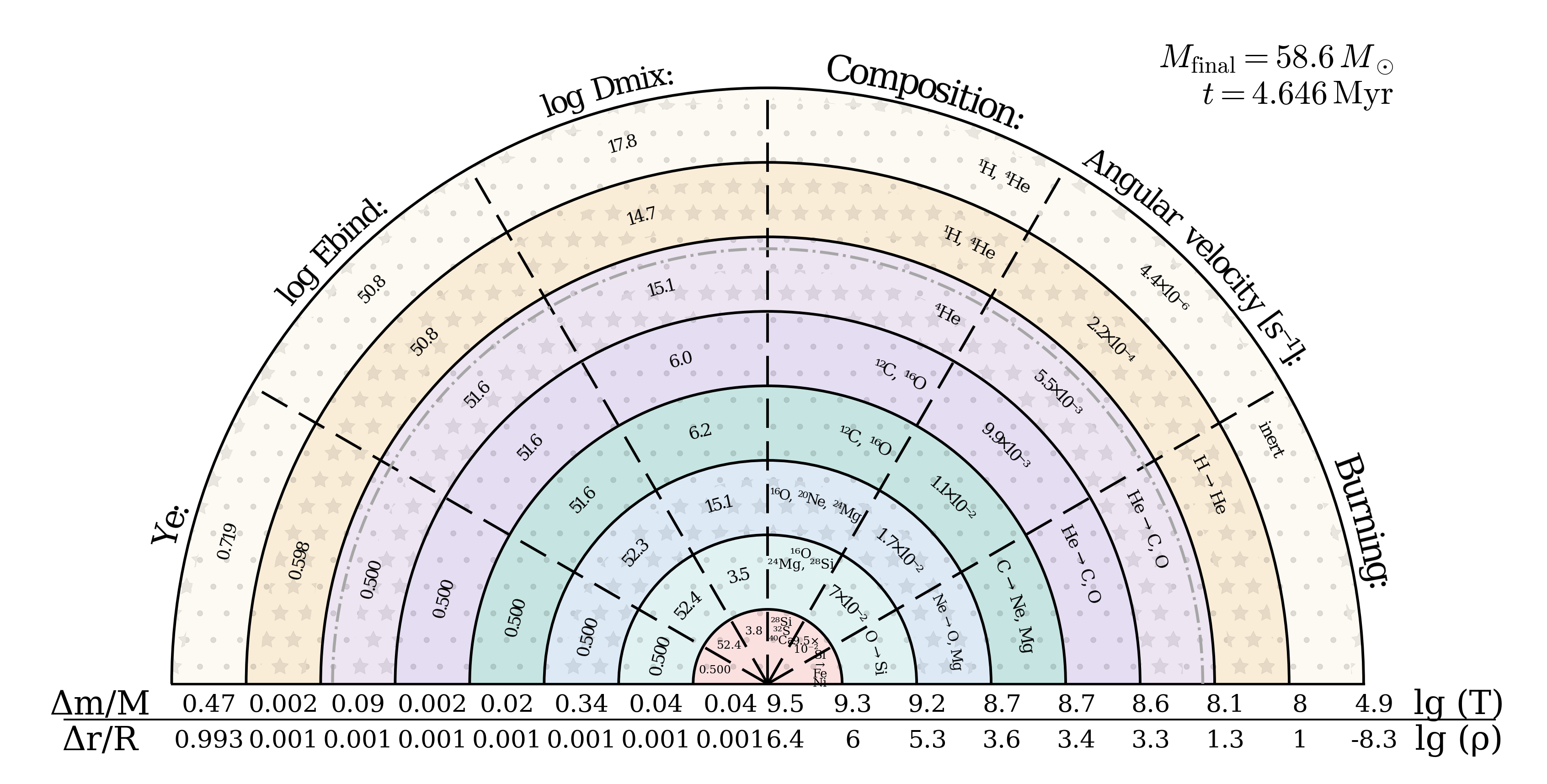}
    \caption{Semicircular shell map of the $60\,M_\odot$ model at the end of core O-burning. Each radial band denotes one chemically distinct shell. The bands are plotted with schematic equal radial spacing, while the physical shell mass and radius fractions are listed below as $\Delta m/M$ and $\Delta r/R$. The lower axis also gives the shell-boundary values of $\log T$ and $\log \rho$. The six angular sectors show, from left to right, the electron fraction $Y_e$, the overlying gravitational binding energy $\log_{10}(E_{\rm bind}/{\rm erg})$, the total mixing coefficient $\log_{10}(D_{\rm mix}/{\rm cm^2\,s^{-1}})$, the dominant composition, the mean angular velocity, and the dominant burning stage. Dot hatching marks radiative regions and star hatching marks convective regions. The gray dash-dotted arc marks the adopted ejection boundary defined by $E_{\rm bind}(>m)=1\,{\rm B}$.}
    \label{fig:onion_120}
\end{figure*}

We use the $60\,M_\odot$ model to illustrate the final structure of the core-collapse branch at the end of core O-burning. Figure~\ref{fig:onion_120} shows how a rapidly-rotating Pop~III star enters the yield calculation, while the dash-dotted arc marks the fiducial $E_{\rm bind}(>m)=1\,{\rm B}$ ejection boundary.
 
The burning and composition slices define the basic shell sequence from center to surface. The dominant layers are Si/S/Ca, O/Mg/Si, O/Ne/Mg, C/O, a very thin H-burning shell, and an inert H/He envelope. The envelope is still dominated by H and He, although some heavier material has been transported outward, and we show only the dominant species in each zone. The processed layers below the envelope contain about $57\%$ of the final mass, while the inert envelope alone contains $\Delta m/M\simeq0.43$. The O/Ne/Mg shell is the largest processed reservoir with $\Delta m/M\simeq0.34$, whereas the active H-burning shell contains only $\Delta m/M\simeq10^{-3}$.

This mass distribution is paired with an even stronger radial compression of the core. The envelope alone spans $\Delta r/R\simeq0.993$, so the advanced-burning shells are confined to only the inner $\sim0.005$ of the stellar radius. The lower boundary values of $\log T$ and $\log \rho$ show the same stratification. From center to surface, $\log T$ falls from about $9.5$ to $4.9$, while $\log \rho$ falls from about $6.4$ to $-8.3$. The final model therefore consists of a hot dense burning core beneath a cool dilute envelope.

The $Y_e$ slice measures the number of protons per baryon. Values above $0.5$ are proton rich, values below $0.5$ are neutron rich, and $Y_e=0.5$ corresponds to nearly equal proton and neutron number. In this model the deep processed layers remain very close to $Y_e=0.500$, so the hydrostatic core is not yet neutron rich at this stage. The H-rich envelope instead reaches $Y_e\simeq0.719$, which reflects its larger proton fraction.

The $\log E_{\rm bind}$ slice gives the overlying gravitational binding energy used in our fiducial yield prescription. It decreases outward from about $10^{52.3}\,{\rm erg}$ in the inner core to about $10^{50.8}\,{\rm erg}$ in the outer layers. Under the $1\,{\rm B}$ criterion, the full H/He envelope and the thin H-burning shell lie on the ejectable side of the cut together with only the outer part of the C/O layer. Most of the O/Ne/Mg, O/Mg/Si, and Si/S/Ca-rich material therefore remains bound. The exact boundary depends on the adopted prescription, but this approach is the most consistent choice for the present grid because the models stop at core O-burning and do not include a collapse calculation.

The $\log D_{\rm mix}$ slice measures the shell-averaged sum of $D_{\rm conv}$, $D_{\rm shear}$, and $D_{\rm eff}$, the convective, shear and effective mixing coefficients, respectively. Its values range from about $10^{3.5}$--$10^{3.8}$ in the deepest shells to about $10^{6.2}$--$10^{17.8}$ farther out, so the local transport rate across a fixed length scale is orders of magnitude larger in the shell system and envelope than in the compact core. The angular-velocity slice shows the same radial decoupling. The innermost shell rotates at $\Omega\simeq9.5\times10^{-2}\,{\rm s^{-1}}$, while the surface envelope is down at only $\sim4.4\times10^{-6}\,{\rm s^{-1}}$. The surface is therefore almost spun down even though the core still rotates rapidly, and the star remains strongly differentially rotating at the end of core O-burning.

The hatching shows that most of the final structure is radiative, while convection is confined to selected shells, and the outermost layers of the envelope. Transport toward the surface therefore depends mainly on rotationally mediated diffusion rather than on deep convective homogenization. Even in a fast rotator, that geometry limits the efficiency with which processed material can be coupled to the outer envelope.

Taken together, these slices show a compact processed core below a massive dilute envelope, with sharp gradients in composition, temperature, density, binding energy, transport efficiency, and rotation. The elemental yields that follow can therefore be read as the integrated contribution of specific shell reservoirs, while the later abundance ratios will show which of those reservoirs dominate the ejecta across the mass grid.

\subsection{Final mass budget and adopted ejecta}
\label{sec:mass_budget}

\begin{figure}
    \centering
    \includegraphics[width=\columnwidth]{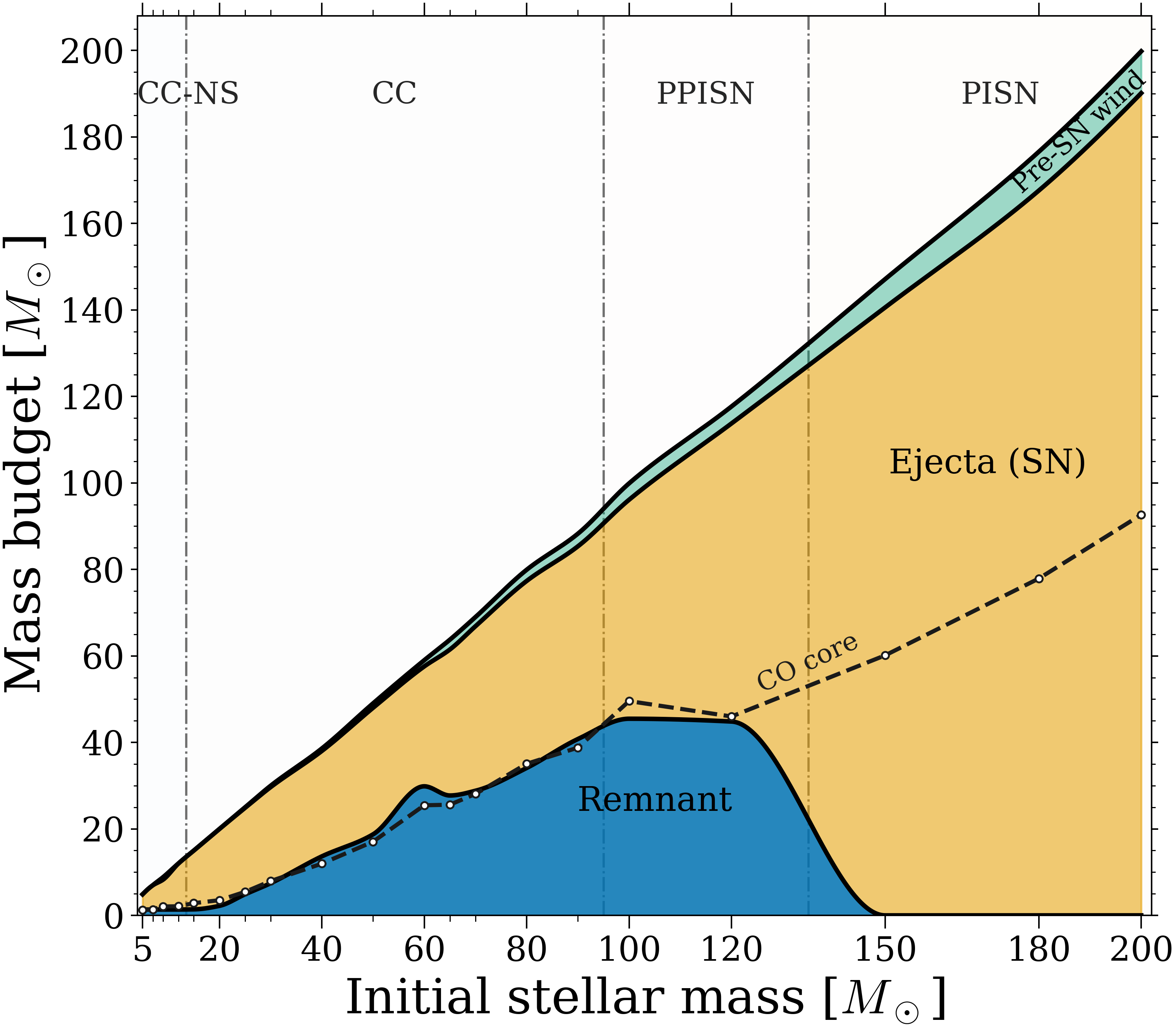}
    \caption{Final mass budget as a function of initial stellar mass. The blue region shows the adopted remnant mass, the gold region shows the ejecta mass, and the green band shows the pre-SN wind mass. The black upper boundary gives the initial stellar mass. The dashed curve marks the CO core mass. Vertical dash-dotted lines separate the adopted CC+NS, CC, PPISN, and PISN regimes.}
    \label{fig:mass_budget}
\end{figure}

Figure~\ref{fig:mass_budget} shows the final mass partition across the grid.  The upper black curve gives the initial mass, while the green band gives the mass lost before the final model, \(M_{\rm wind}=M_{\rm ini}-M_{\rm final}\).  The blue region is the adopted remnant mass, and the gold region is the material exterior to this remnant, \(M_{\rm ej}=M_{\rm final}-M_{\rm rem}\). The dashed curve marks the CO core mass.

At the low-mass end of the core-collapse branch, the adopted remnant is set by the neutron-star floor.  The \(5\), \(7\), \(9\), and \(12\,M_\odot\) models all have \(M_{\rm rem}=1.35\,M_\odot\), while the ejecta mass rises from \(3.50\) to \(10.64\,M_\odot\).  Their CO cores remain small, with \(M_{\rm CO}=1.24\)--\(2.11\,M_\odot\), so the binding-energy criterion does not require a massive fallback remnant.  These models therefore eject most of their final H/He-rich envelopes, but they contribute little from the deeper processed core.

The main core-collapse sequence begins once the CO core becomes more massive and more tightly bound.  The adopted remnant mass rises from \(2.20\,M_\odot\) at \(20\,M_\odot\) to \(13.64\,M_\odot\) at \(40\,M_\odot\), \(29.85\,M_\odot\) at \(60\,M_\odot\), and \(40.77\,M_\odot\) at \(90\,M_\odot\). Over the same interval, the CO core mass increases from \(3.51\) to \(38.79\,M_\odot\). The remnant and CO core curves therefore track each other closely through most of the core-collapse regime. This occurs because the compact processed core dominates the binding energy barrier, while the extended outer layers remain easier to eject. The consequence is that the deepest O, Si, and Fe group producing layers are mostly retained, while the adopted ejecta are weighted toward the envelope and the outer processed shells.

The ejecta mass in the core-collapse branch grows more slowly than the initial mass because the remnant also grows. It increases from \(17.77\,M_\odot\) at \(20\,M_\odot\) to \(24.37\,M_\odot\) at \(40\,M_\odot\), \(27.72\,M_\odot\) at \(60\,M_\odot\), and \(44.56\,M_\odot\) at \(90\,M_\odot\). The non-linear growth reflects the changing position of the binding-energy cut relative to the shell structure. This is the main reason that the elemental yields need not scale monotonically with initial mass, even when the final mass itself increases.

The PPISN models occupy an intermediate branch between ordinary core collapse and complete disruption. The \(100\,M_\odot\) and \(120\,M_\odot\) models leave similar remnants, \(45.47\) and \(44.82\,M_\odot\), while their ejecta masses increase from \(50.64\) to \(68.93\,M_\odot\). This produces the plateau-like remnant mass around \(100\)--\(120\,M_\odot\). Physically, pulsational mass ejection removes part of the star, but the final collapse still leaves a compact remnant. The yield consequence is that the ejecta become more massive than in the lower core-collapse models, but the innermost processed material is still not fully released.

The PISN branch is qualitatively different. The adopted remnant mass is zero, so the \(150\), \(180\), and \(200\,M_\odot\) models eject their full final masses of \(140.59\), \(167.60\), and \(190.14\,M_\odot\). Once the pair-instability disruption occurs, the entire final star is unbound. This releases both the envelope and the processed core, and it therefore marks the regime where the elemental yields should increase most strongly.

The pre-SN wind is a secondary term in the total mass budget, but it is still visible at high mass. The wind loss is \(0.15\,M_\odot\) at \(5\,M_\odot\), \(1.99\,M_\odot\) at \(40\,M_\odot\), \(6.24\,M_\odot\) at \(120\,M_\odot\), and \(12.40\,M_\odot\) at \(180\,M_\odot\). These losses remain only a few percent of the initial mass over most of the grid.  They therefore do not control the total ejecta mass, but they are important because they carry the surface composition established by rotational mixing before the final fate.

The key result from Fig.~\ref{fig:mass_budget} is that the final yields are controlled primarily by the location of the remnant boundary relative to the CO core. In the core-collapse and PPISN regimes, the mass cut lies close to the compact processed core and much of that material is retained. In the PISN regime, the entire final star is ejected. This transition sets the expected change in the elemental-yield pattern discussed next.

\subsection{Gross elemental ejecta masses}
\label{sec:individual_elemental_yields}

\begin{figure*}
    \centering
    \includegraphics[width=\textwidth]{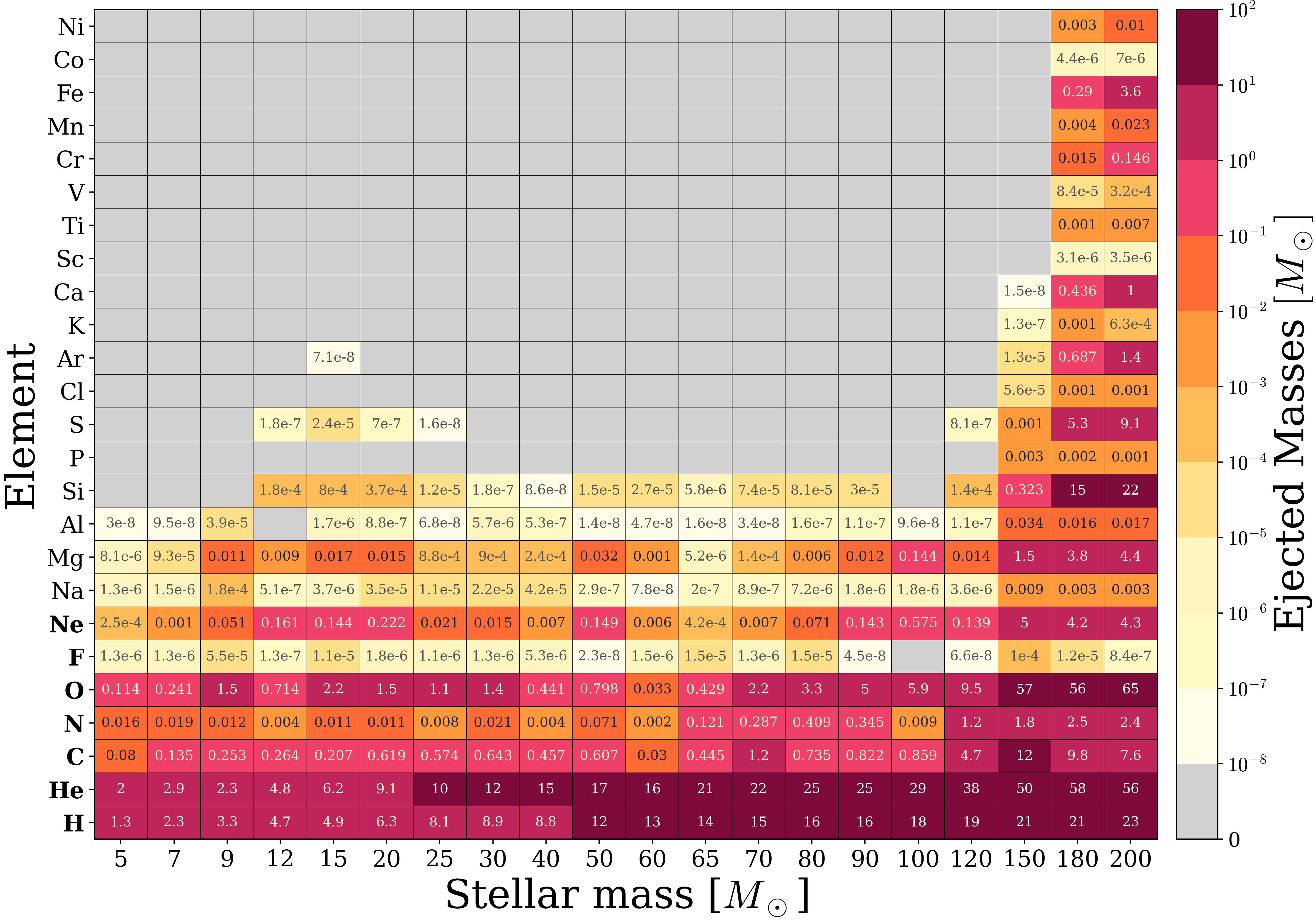}
    \caption{Gross elemental masses in the adopted ejecta of the $Z=0$, $\upsilon_{\rm ini}/\upsilon_{\rm crit}=0.7$ grid, including pre-SN wind material where present. CNO represents the principal hydrostatic nucleosynthetic result of the present models. The CC and PPISN values for elements heavier than Ne are stage-dependent final-model reservoirs rather than hydrodynamic explosion yields. Grey cells with crosses mark values below $10^{-8}\,M_\odot$.}
    \label{fig:element_yields}
\end{figure*}

Figure~\ref{fig:element_yields} translates the mass partition in Fig.~\ref{fig:mass_budget} into elemental ejecta masses. The central result is that rotation enhances CNO-rich ejectable material, while the remnant boundary controls how much of the deeper alpha-rich core is released. This follows directly from the shell geometry discussed above for the $120\,M_\odot$ model. The H/He envelope and outer processed material lie on the ejectable side of the cut, whereas much of the compact O/Ne/Mg and Si/S/Ca-rich core remains bound in retained-core models. The neutron-star-forming core-collapse (NS--CC) and CC models therefore produce H/He-diluted CNO enrichment with limited deep-core contribution. The PPISN models eject more total mass but still retain a compact remnant. The PISN models remove that filter because the adopted remnant mass is zero. All values quoted below are gross adopted ejecta masses and include pre-SN wind material where present.

H and He set the dilution scale for all heavier elements. The H yield increases overall from $1.33\,M_\odot$ at $5\,M_\odot$ to $15.79\,M_\odot$ at $90\,M_\odot$. It then reaches $18.21$--$19.28\,M_\odot$ in the PPISN models and $20.81$--$22.91\,M_\odot$ in the PISN branch. He grows from $2.01\,M_\odot$ at $5\,M_\odot$ to $38.00\,M_\odot$ at $120\,M_\odot$, then reaches $49.59$--$57.56\,M_\odot$ after complete disruption. These large H/He masses show that the adopted ejecta are never pure core material. The wind is usually secondary in total mass, but it is chemically important because it carries the rotation-enriched surface composition before the final fate is applied.

C, N, and O carry the main rotational signal in the retained-core regimes, so they are the most important products for early CNO enrichment. The C yield is already $7.97\times10^{-2}$--$2.64\times10^{-1}\,M_\odot$ in the NS--CC models. It reaches $6.19\times10^{-1}\,M_\odot$ at $20\,M_\odot$, falls to $3.00\times10^{-2}\,M_\odot$ at $60\,M_\odot$, and recovers to $1.20\,M_\odot$ at $70\,M_\odot$. This non-monotonic behavior reflects where the adopted cut intersects the C/O reservoir. The $120\,M_\odot$ PPISN model ejects $4.69\,M_\odot$ of C, while the PISN branch ejects $11.95$, $9.80$, and $7.57\,M_\odot$. The decline across the PISN sequence occurs while O and Si increase, indicating more complete processing of C into heavier alpha nuclei. C is a tracer of He-burning material that escapes full conversion into oxygen.

N is the clearest primary product of fast rotation in the grid. Since the initial metallicity is zero, the ejected N must be made from newly synthesized C and O. The NS--CC models eject $3.75\times10^{-3}$--$1.86\times10^{-2}\,M_\odot$ of N, but the retained-core sequence spans more than two orders of magnitude. The yield is only $1.82\times10^{-3}\,M_\odot$ at $60\,M_\odot$ and reaches $4.09\times10^{-1}\,M_\odot$ at $80\,M_\odot$. The contrast is even sharper between the PPISN models, from $8.89\times10^{-3}\,M_\odot$ at $100\,M_\odot$ to $1.24\,M_\odot$ at $120\,M_\odot$. This sensitivity shows that N is controlled by rotational coupling between He-burning products and H-burning layers. The PISN models eject $1.81$--$2.53\,M_\odot$ of N, making these fast rotators a prompt source of primary N.

O dominates the heavy-element mass once O-rich layers are exposed. The NS--CC branch peaks at $1.51\,M_\odot$ in the $9\,M_\odot$ model because part of its O-rich region lies outside the fixed neutron-star remnant. In the CC branch the yield is $2.15\,M_\odot$ at $15\,M_\odot$, drops to $3.26\times10^{-2}\,M_\odot$ at $60\,M_\odot$, and rises again to $5.01\,M_\odot$ at $90\,M_\odot$. The PPISN models eject $5.89$--$9.51\,M_\odot$ of O. The PISN branch then marks the largest transition in the figure, with $57.03$--$65.38\,M_\odot$ of O ejected. O therefore traces the release of the He-burning core and O-rich shells. It remains modest when those layers are retained and becomes the dominant metal reservoir when the whole star is disrupted.

Ne and Mg mark partial access to the O/Ne/Mg region. Ne reaches $2.22\times10^{-1}\,M_\odot$ at $20\,M_\odot$, falls to $6.24\times10^{-3}\,M_\odot$ at $60\,M_\odot$, and rises to $5.75\times10^{-1}\,M_\odot$ in the $100\,M_\odot$ PPISN model. The PISN models eject much larger Ne masses of $4.22$--$5.01\,M_\odot$. Mg is usually below $3.2\times10^{-2}\,M_\odot$ through the CC branch, but reaches $1.44\times10^{-1}\,M_\odot$ at $100\,M_\odot$ and $1.54$--$4.41\,M_\odot$ in the PISN branch. These jumps show that much of the O/Ne/Mg reservoir lies near the retained-core mass boundary. Ne and Mg therefore mark the first access to material deeper than the CNO-rich layers. Since the calculations presented here terminate during core O burning, the reported yields are representative of this evolutionary stage. For the lower-mass models, subsequent core Si burning would alter the abundances of elements beyond Ne before core collapse, so the values reported here should not be interpreted as their final pre-supernova yields.

The odd-Z elements remain secondary tracers rather than major contributors to the ejecta mass. Na stays below $10^{-3}\,M_\odot$ before the PISN branch and reaches only $9.15\times10^{-3}\,M_\odot$ at $150\,M_\odot$. Al is below $4\times10^{-5}\,M_\odot$ before complete disruption, then rises to $3.37\times10^{-2}\,M_\odot$ at $150\,M_\odot$. F remains below $1.02\times10^{-4}\,M_\odot$ across the grid. Its connection to primary N seeding makes it physically interesting, but its small mass prevents it from affecting the total enrichment budget. P, Cl, and K appear mainly in the PISN branch and remain at the $10^{-5}$--$10^{-3}\,M_\odot$ level. The weakness of these species reflects the limited neutron excess in most of the ejectable zero-metallicity material.

The deeper $\alpha$-chain elements provide the clearest signature of complete disruption. 
Si remains below $8.0\times10^{-4}\,M_\odot$ through the CC branch and is only $1.41\times10^{-4}\,M_\odot$ at $120\,M_\odot$. It then rises to $3.23\times10^{-1}\,M_\odot$ at $150\,M_\odot$ and $21.74\,M_\odot$ at $200\,M_\odot$. 
S follows the same retained-core suppression, staying below $2.39\times10^{-5}\,M_\odot$ in the CC branch and reaching only $8.12\times10^{-7}\,M_\odot$ at $120\,M_\odot$. The PISN branch ejects $1.19\times10^{-3}$--$9.06\,M_\odot$ of S. Ar and Ca are concentrated in the two highest PISN models, with Ar reaching $6.87\times10^{-1}$--$1.36\,M_\odot$ and Ca reaching $4.36\times10^{-1}$--$1.04\,M_\odot$. These species sit deeper than the CNO and O/Ne/Mg reservoirs. Their sudden appearance shows when the full processed core is no longer hidden by remnant retention.

We do not use the Fe-group cells to draw conclusions about the {\gva} core-collapse or PPISN models in this subsection. The models stop at core O-burning, and the CC and PPISN entries are final-model reservoirs rather than hydrodynamic explosion yields. The robust abundance result is instead the separation between retained-core and full-disruption enrichment. 
The NS--CC and CC models can seed primordial gas with H/He-diluted CNO material while contributing little Si, S, Ar, or Ca. The PPISN models increase the total ejecta mass but preserve the same filtering by a compact remnant. The PISN models remove that filter and eject very large O/Ne/Mg and Si/S/Ca masses. Thus, rotation prepares the CNO-rich layers, while the final fate determines whether the compact alpha-rich core remains hidden or is released.

\section{Discussion}
\label{sec:Discussion}
\subsection{Fast-rotator yields and high-redshift abundance anomalies}
\label{sec:abundance_ratio_diagnostics}

\begin{figure*}
    \centering
    \includegraphics[width=\textwidth]{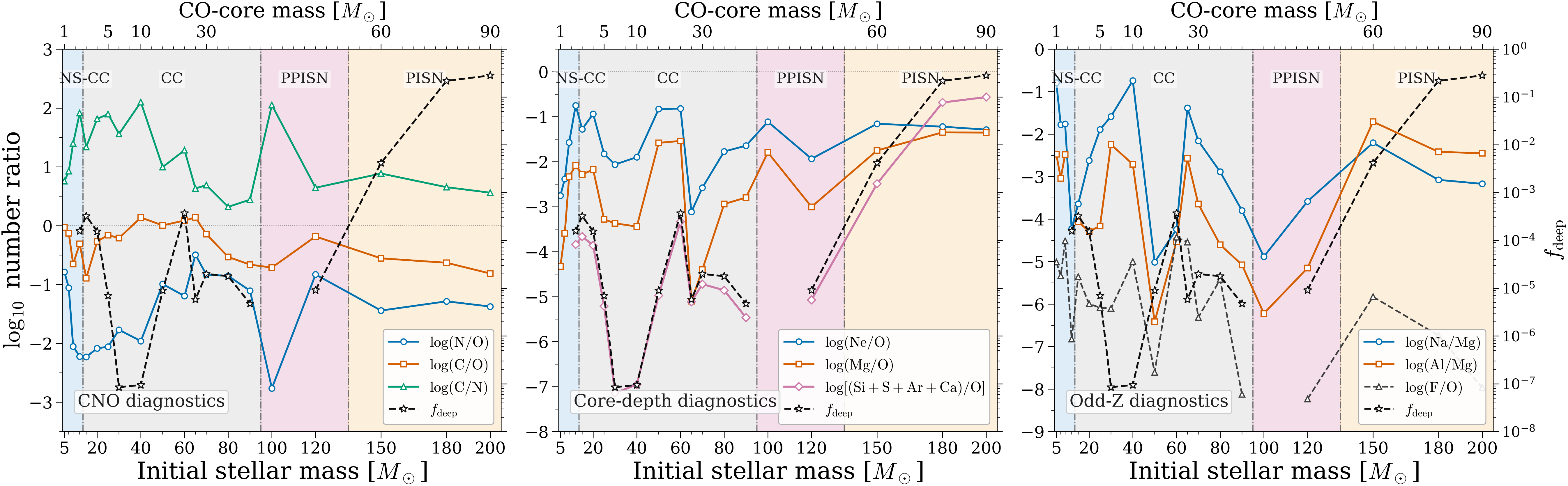}
    \caption{
    Number-fraction abundance ratios for the $Z=0$, $\upsilon_{\rm ini}/\upsilon_{\rm crit}=0.7$ grid. 
    The left panel shows CNO ratios, the middle panel shows ratios that trace how deeply the ejecta sample the processed core, and the right panel shows selected odd-Z ratios. 
    The dashed black curve in each panel is 
    $f_{\rm deep}=(M_{\rm Si}+M_{\rm S}+M_{\rm Ar}+M_{\rm Ca})/
    (M_{\rm C}+M_{\rm N}+M_{\rm O}+M_{\rm Ne}+M_{\rm Mg}+M_{\rm Si}+M_{\rm S}+M_{\rm Ar}+M_{\rm Ca})$. 
    Shaded regions mark the NS--CC, CC, PPISN, and PISN regimes.
    }
    \label{fig:abundance_ratio_diagnostics}
\end{figure*}

Figure~\ref{fig:abundance_ratio_diagnostics} depicts number-fraction ratios in the adopted ejecta. 
The ratios include the terminal ejecta and the pre-SN wind contribution. 
We focus on CNO, alpha-element, and odd-Z diagnostics, while Fe-based ratios are omitted because the CC and PPISN models are not hydrodynamic explosion yields. 
The dashed black curve in each panel shows $f_{\rm deep}$, the fraction of the main metal ejecta carried by Si, S, Ar, and Ca.

The left panel of Fig.~\ref{fig:abundance_ratio_diagnostics} shows that N/O is enhanced only in selected retained-core models. 
The maximum value is $\log({\rm N/O})=-0.49$ at $65\,M_\odot$, while the $100\,M_\odot$ PPISN model drops to $-2.76$. 
The PISN models cluster at lower values, with $\log({\rm N/O})=-1.44$ to $-1.29$. 
This behavior reflects where primary N is made and where oxygen is released. 
Rotational mixing brings newly synthesized C and O into H-burning layers, where CNO cycling builds primary N. 
Once the whole star is disrupted, the large O-rich core dominates the ejecta and drives N/O back down.

This comparison is restrictive for the N-rich JWST sources. 
GN-z11 requires $\log({\rm N/O})>-0.25$, CEERS-1019 has $\log({\rm N/O})=-0.18\pm0.11$, and GS~3073 reaches $\log({\rm N/O})=0.46^{+0.12}_{-0.09}$ \citep{Cameron2023,Marques2023,Ji2024, Ji2026}. 
Our $5$--$200\,M_\odot$ grid remains below these systems even with $\upsilon_{\rm ini}/\upsilon_{\rm crit}=0.7$. 
The closer comparisons are more moderate N emitters such as RXCJ2248-ID, with $\log({\rm N/O})=-0.39^{+0.11}_{-0.10}$, or GN-z9p4, whose high-ionization nitrogen ratio is near $-0.59$ \citep{Topping2024,Schaerer2024}. 
Ordinary massive fast rotators therefore provide a primary N source, but the most extreme N/O systems likely require a different weighting of the enrichment channel or much higher stellar masses. 
This agrees with models in which $10^3$--$10^4\,M_\odot$ Pop~III stars reproduce the N/O, C/O, and Ne/O ratios of GS~3073 \citep{Nandal2025}.

C gives the complementary constraint. 
The retained-core models can be C rich, with $\log({\rm C/O})$ reaching $0.14$ near $40$--$65\,M_\odot$. 
The PISN branch is more O dominated and gives $\log({\rm C/O})=-0.55$ to $-0.81$. 
Several high-redshift measurements lie in this O-rich range. 
CEERS-1019 has $\log({\rm C/O})=-0.75\pm0.11$, GN-z11 has $\log({\rm C/O})>-0.78$, and an early JWST $z>8$ abundance study found $\log({\rm C/O})=-0.83\pm0.38$ \citep{Marques2023,Cameron2023,Arellano2022}. 
The EXCELS galaxies at $z\simeq5$ give similarly low values of $-0.82\pm0.22$ and $-1.02\pm0.22$, while MACS0647-JD is more C rich with $\log({\rm C/O})=-0.44^{+0.06}_{-0.07}$ at $z=10.17$ \citep{Arellano2025,Hsiao2025}. 
The C/N ratio then isolates the missing piece. 
Our grid has a minimum $\log({\rm C/N})=0.32$, whereas CEERS-1019 and GS~3073 imply $\log({\rm C/N})\simeq-0.57$ and $-0.76$ \citep{Marques2023,Ji2024}. 
The most N-dominated galaxies therefore appear to require more complete conversion of C into N than these ordinary massive-star ejecta provide.

In \citet{Martinez2025} it is presented the revised nebular abundances for several N-rich high-redshift galaxies, including GN-z11, GS 3073 and CEERS-1019. They use a multi-phase ionization model, that explicitly accounts for the density of the high-ionization zone and applies ionization correction factors absent from earlier single-zone determinations. The literature values we adopt above instead, follow the original analyses \citep{Cameron2023, MarquesChaves2024, Ji2024}, which lacked this density correction treatment.

Under the \citet{Martinez2025} correction, $12+log(O/H)$ rises by $\sim 0.5$ dex for GN-z11 and GS 3073, while $log(N/O)$ us sensitive to the adopted ionic tracer. The $N^{+2}/O^{+2}$-based ratio brings CEERS-1019 to $log(N/O)=-0.54$, near our retained-core maximum of $-0.49$, whereas the $N^{+3}/O^{+3}$-based ratio remains high for all three systems ($log(N/O)=0.17-1.13$). Using the \citet{Martinez2025} values, CEERS-1019 approaches the upper end of our retained-core models, whereas GN-z11 and GS 3073 remain more N-enhanced than our $5$--$200 \, M_\odot$ rapidly-rotating Pop III grid can reproduce. Our main conclusion therefore remains unchanged here, as we find that ordinary fast rotators can produce substantial primary N-enrichment, but the most extreme high-redshift N/O ratios likely require either selective enrichment or an additional very massive stellar channel.

The middle panel of Fig.~\ref{fig:abundance_ratio_diagnostics} tests whether the ejecta include material deeper than the CNO layers. 
Ne/O and Mg/O trace partial access to the O/Ne/Mg shell, while $f_{\rm deep}$ tracks the Si/S/Ar/Ca-rich material. 
The retained-core models span $\log({\rm Ne/O})=-3.11$ to $-0.75$, while observed values in CEERS-1019, EXCELS, and GN-z9p4 lie near $-0.7$ \citep{Schaerer2024, Arellano2025}. 
Only the upper edge of our Ne/O range approaches those systems. 
At the same time, $f_{\rm deep}$ remains tiny through most retained-core models and rises to $0.22$--$0.28$ in the $180$--$200\,M_\odot$ PISN models. 
The clearest signature of retained-core enrichment is therefore CNO production without a matching rise in Si/S/Ar/Ca.

The right panel of Fig.~\ref{fig:abundance_ratio_diagnostics} gives secondary odd-Z checks. 
Na/Mg and Al/Mg vary by several dex because they depend on neutron excess and shell processing. 
F/O is still more fragile because F requires primary N seeding but is easily destroyed at high temperature. 
These ratios are unlikely to identify the enrichment channel alone, but they can test consistency once CNO and alpha-element ratios are measured. 
Taken together, the three panels show that rapidly-rotating \(5\)--\(200\,M_\odot\) Pop~III models can produce CNO-rich retained-core ejecta, but not the most N-dominated JWST abundance patterns. 
Those systems point toward selective enrichment by the strongest N-producing models, or toward a separate very massive to supermassive stellar channel.

\subsection{Population-averaged abundance ratios}
\label{sec:imf_weighted_ratios}

\begin{figure}
    \centering
    \includegraphics[width=\columnwidth]{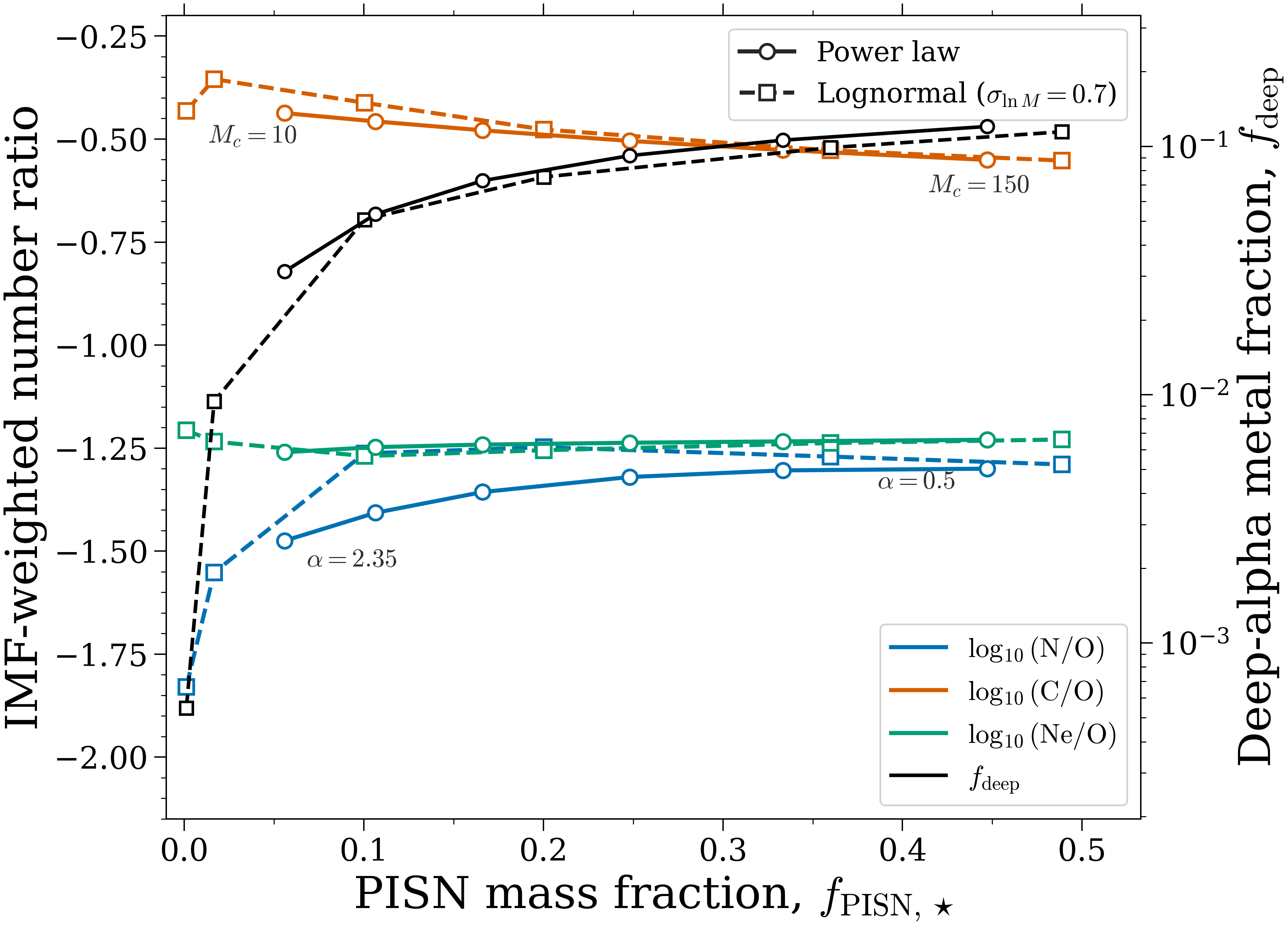}
    \caption{IMF-weighted abundance ratios over the computed $5$--$200\,M_\odot$ rapidly-rotating Pop~III grid. Solid curves show power-law weights, $dN/dM\propto M^{-\alpha}$, and dashed curves show lognormal weights with $\sigma_{\ln M}=0.7$. The x-axis gives the stellar mass fraction assigned to PISN progenitor bins. Coloured curves show IMF-weighted number ratios. The black curve shows $f_{\rm deep}$, the fraction of the main metal ejecta in Si, S, Ar, and Ca.}
    \label{fig:imf_weighted_ratios}
\end{figure}

The previous section compared individual models to observed abundance patterns. 
We now ask how the same yields behave after weighting the computed $5$--$200\,M_\odot$ grid by idealized Pop~III IMFs. 
Figure~\ref{fig:imf_weighted_ratios} shows two families of prompt-enrichment weights: power laws, $dN/dM\propto M^{-\alpha}$, and lognormal distributions with characteristic mass $M_c$. 
These are weights over the computed grid rather than a complete IMF including lower-mass stars. 
The horizontal axis gives $f_{\rm PISN,\star}$, the fraction of the initial stellar mass assigned to PISN progenitor bins.

The IMF-weighted CNO ratios reinforce the conclusion from Fig.~\ref{fig:abundance_ratio_diagnostics}. 
A more top-heavy population does not automatically produce higher N/O. 
For the power-law sequence, $\log({\rm N/O})$ rises only from $-1.47$ at $\alpha=2.35$ to $-1.30$ at $\alpha=0.5$. 
The lognormal sequence reaches a similar value, $\log({\rm N/O})=-1.29$, at $M_c=150\,M_\odot$. 
These values remain well below the most N-dominated JWST systems discussed above. 
The reason is that increasing the PISN weight adds a large O reservoir along with any additional N. 
Top-heavy weighting therefore makes the population more O rich before it makes it N dominated.

C and Ne respond more weakly to the IMF shape. 
Across the power-law sequence, $\log({\rm C/O})$ changes from $-0.44$ to $-0.55$, while $\log({\rm Ne/O})$ stays close to $-1.25$. 
The lognormal models span a similar range, with $\log({\rm C/O})=-0.43$ at $M_c=10\,M_\odot$ and $-0.55$ at $M_c=150\,M_\odot$. 
Thus, IMF weighting mostly shifts the mixture toward O-rich full-disruption ejecta. 
It does not produce the low C/N and high N/O combination required by the most extreme high-redshift nitrogen sources.

The strongest population-level trend is the rise of $f_{\rm deep}$. 
For the lognormal sequence, $f_{\rm deep}$ increases from $5.5\times10^{-4}$ at $M_c=10\,M_\odot$ to $0.11$ at $M_c=150\,M_\odot$. 
For the power-law sequence, it rises from $0.031$ at $\alpha=2.35$ to $0.12$ at $\alpha=0.5$. 
A more top-heavy population therefore increases the contribution of Si/S/Ar/Ca-rich material. 
The integrated ejecta move from a retained-core-like CNO pattern toward a PISN-like deep-alpha pattern.

The IMF-weighted result sharpens the interpretation of the single-star ratios. 
Fast rotation supplies primary N, but IMF weighting over $5$--$200\,M_\odot$ mainly controls how much O-rich and deep-alpha material is added. 
The limiting factor for the most N-rich JWST systems is therefore the balance between primary N-producing retained-core ejecta and oxygen-rich full-disruption ejecta.

\subsection{Sensitivity to the adopted remnant-mass prescription}
\label{subsec:remnant_prescription_test}

Our fiducial pre-SN yields are computed by placing the mass cut with the binding-energy criterion described above. To test whether this choice drives our conclusions, we repeated the yield calculation with two commonly used remnant-mass prescriptions. First, we used the CO core-based prescription of \citet{Maeder1992}. Second, we applied a Patton--Sukhbold/Ertl-style explodability treatment, following the structure-based approach of \citet{Patton2020} and \citet{Ertl2016}. In each case, only the adopted remnant mass was changed. The ejecta and yields were then recomputed by integrating the final abundance profiles exterior to the corresponding mass cut.

The Maeder prescription gives substantially smaller remnants and therefore larger ejecta masses.  This is expected, since it represents an older and more ejective treatment of black-hole formation and fallback. The Patton--Sukhbold/Ertl comparison is more relevant for assessing our fiducial method. Where the published Patton--Sukhbold CO core grid can be used directly, the individual mass cuts shift from model to model, but the resulting ejecta masses remain within a factor of \(0.83\)--\(1.25\) of our binding-energy values over \(15\)--\(30\,M_\odot\). For the more massive models, whose CO cores lie outside the published grid, we do not treat the calculation as a direct PS20 prediction. Instead, we adopt a conservative failed-collapse continuation with \(M_{\rm rem}=M_{\rm He}\). Over \(40\)--\(120\,M_\odot\), this changes the remnant mass by at most \(2.7\,M_\odot\), and gives ejecta masses that are \(0.89\)--\(0.99\) times the fiducial values.

We therefore conclude that our main yield estimates are not the result of an anomalously chosen mass cut.  The Maeder prescription brackets a more extreme, older alternative, while the Patton--Sukhbold/Ertl-style treatment supports the same qualitative outcome as our fiducial calculation.  The massive models retain large compact remnants, and the pre-SN ejecta remain within the same order of magnitude.

\subsection{Comparison with previous Pop~III yield calculations}
\label{sec:comparison_previous_yields}

Non-rotating Pop~III explosion grids provide the baseline for judging what rotation changes. 
The zero-metallicity calculations of \citet{Chieffi2004} and \citet{Limongi2012} cover $13$--$35\,M_\odot$ and $13$--$80\,M_\odot$, respectively, and follow explosive nucleosynthesis through the Fe group. 
Our models stop at the end of core O-burning, so we do not compare Fe-group yields here. 
The useful comparison is instead the behavior of CNO and alpha-rich material before the explosion physics is added. 
In our overlapping CC mass range, $\log({\rm C/O})$ varies from $-0.89$ at $15\,M_\odot$ to $0.14$ at $65\,M_\odot$, then falls to $-0.53$ at $80\,M_\odot$. 
This non-monotonic behavior differs from standard non-rotating explosion grids, where C/O is mainly set by progenitor mass and explosion/mass-cut choices. 
It reflects the combined effect of fast rotational mixing and the adopted remnant boundary.

The comparison with rotating Pop~III grids is more direct. 
In the $\upsilon_{\rm ini}/\upsilon_{\rm crit}=0.4$ {\gva} grid, the rotating $12$ and $15\,M_\odot$ models produce $^{14}{\rm N}$ masses of order $8\times10^{-3}$ and $1.2\times10^{-2}\,M_\odot$, far above their non-rotating counterparts \citep{Murphy2021}. 
Our adopted ejecta give comparable low-mass values, with $M_{\rm N}=3.75\times10^{-3}$ and $1.11\times10^{-2}\,M_\odot$ at $12$ and $15\,M_\odot$. 
The difference appears at higher mass. 
With faster initial rotation and later evolutionary endpoints, our grid reaches $M_{\rm N}=0.29$--$0.41\,M_\odot$ at $70$--$80\,M_\odot$ and $1.24\,M_\odot$ at $120\,M_\odot$. 
This confirms the same primary N channel but extends it into a regime where final fate strongly controls the observable source ratios.

The closest same-rotation comparison is the $\upsilon_{\rm ini}/\upsilon_{\rm crit}=0.7$ grid of \citet{Tsiatsiou2024}. 
Those models show that fast rotation does not require chemically homogeneous evolution to make primary N. 
Instead, the key structure is an extended H-burning shell that receives C and O from He-burning layers. 
Our models follow the same physical channel, but the yield interpretation is different because we use the final adopted ejecta and extend the comparison to the $150$--$200\,M_\odot$ PISN branch. 
Selected compact-remnant models reach $\log({\rm N/O})=-0.49$ and $\log({\rm C/O})=0.14$. 
The PISN models then return to lower $\log({\rm N/O})=-1.44$ to $-1.29$ because the full O-rich core is released.

Rapidly rotating Pop~III models have also been calculated with KEPLER by \citet{Jeena2023}, who considered $20$--$35\,M_\odot$ stars with $\upsilon_{\rm ini}/\upsilon_{\rm crit}=0.4$--$0.7$. Those models employ diffusive rotational mixing with magnetic angular-momentum transport and parameterized piston and mixing--fallback explosions. They overlap with part of our rotation range but do not reproduce the broad $5$--$200\,M_\odot$ mass coverage or the GENEC transport framework considered here.

This final-fate assumption also explains the difference from recent high-redshift fits. 
Rapidly-rotating Pop~III models sampled with a top-heavy IMF can match GN-z11 and CEERS-1019-like abundances with $\log({\rm N/O})=-0.38$, $\log({\rm C/O})=-0.22$, and $12+\log({\rm O/H})=7.82$ after dilution by factors $f=20$--100 \citep{Nandal2024f}. 
Our individual models approach that N/O value only near the compact-remnant peak. 
After IMF weighting over the full $5$--$200\,M_\odot$ grid, the power-law sequence gives only $\log({\rm N/O})=-1.47$ to $-1.30$. 
A top-heavy weighting therefore adds O and deep-alpha material faster than it builds a N-dominated population.

The behavior of the PISN branch is consistent with classic very massive Pop~III calculations. 
\citet{Heger2002} showed that metal-free PISNe produce strong even-$Z$ products and a pronounced odd-even pattern. 
Our full-disruption models follow the same trend, ejecting $57$--$65\,M_\odot$ of O and raising $f_{\rm deep}$ to $0.22$--$0.28$ at $180$--$200\,M_\odot$. 
The odd-$Z$ species remain secondary compared with the even-$Z$ alpha products. 
At lower mass, the rotating explosion models of \citet{Roberti2024} provide a complementary benchmark because they follow $15$ and $25\,M_\odot$ Pop~III and extremely metal-poor stars through collapse with a large nuclear network. 
Our calculation covers a broader mass range and isolates the pre-explosion effect of the adopted remnant boundary. 
The main result is therefore that rotation enhances CNO-rich ejecta, while compact-remnant formation or complete disruption decides whether the deep alpha-rich core is hidden or released.

\begin{table*}[h]
    \caption{Gross adopted ejecta masses for selected isotopes} for selected isotopes as a function of initial stellar mass ($M_\odot$) over the mass range $5$--$50\,M_\odot$. Yields below $10^{-10}\,M_\odot$ are treated as zero. For CC and PPISN models, species heavier than Ne are stage-dependent reservoirs and not final explosive-nucleosynthesis yields.
    \label{tab:yields_lowmass}
    \scriptsize                 
    \setlength{\tabcolsep}{6pt}    
    \renewcommand{\arraystretch}{1}      
    \resizebox{\textwidth}{!}{
    \hspace*{-0.17\textwidth}
    \begin{tabular}{|l|cccccccccc|}
        \hline
        Isotope & \multicolumn{10}{c|}{Stellar Mass ($M_\odot$)} \\
        \cline{2-11} & $5$ & $7$ & $9$ & $12$ & $15$ & $20$ & $25$ & $30$ & $40$ & $50$ \\
        \hline
        $^{1}$H & $1.33$ & $2.33$ & $3.33$ & $4.69$ & $4.90$ & $6.25$ & $8.11$ & $8.91$ & $8.82$ & $12.1$ \\
        $^{3}$He & $3.0\times10^{-5}$ & $6.4\times10^{-4}$ & $2.9\times10^{-4}$ & $1.9\times10^{-3}$ & $5.1\times10^{-3}$ & $1.1\times10^{-3}$ & $1.7\times10^{-3}$ & $5.8\times10^{-5}$ & $3.3\times10^{-5}$ & $4.8\times10^{-3}$ \\
        $^{4}$He & $2.01$ & $2.92$ & $2.32$ & $4.83$ & $6.15$ & $9.14$ & $10.3$ & $11.6$ & $15.2$ & $16.6$ \\
        $^{12}$C & $7.9\times10^{-2}$ & $1.3\times10^{-1}$ & $2.5\times10^{-1}$ & $2.6\times10^{-1}$ & $1.9\times10^{-1}$ & $6.0\times10^{-1}$ & $5.6\times10^{-1}$ & $6.0\times10^{-1}$ & $4.5\times10^{-1}$ & $6.0\times10^{-1}$ \\
        $^{13}$C & $1.1\times10^{-3}$ & $1.7\times10^{-3}$ & $1.5\times10^{-3}$ & $4.2\times10^{-3}$ & $1.2\times10^{-2}$ & $2.0\times10^{-2}$ & $7.0\times10^{-3}$ & $2.9\times10^{-2}$ & $1.6\times10^{-4}$ & $5.4\times10^{-3}$ \\
        $^{14}$C & $3.6\times10^{-6}$ & $7.8\times10^{-5}$ & $4.4\times10^{-4}$ & $4.0\times10^{-3}$ & $5.0\times10^{-4}$ & $2.5\times10^{-3}$ & $3.1\times10^{-3}$ & $1.2\times10^{-2}$ & $3.9\times10^{-3}$ & $7.1\times10^{-4}$ \\
        $^{14}$N & $1.6\times10^{-2}$ & $1.8\times10^{-2}$ & $1.1\times10^{-2}$ & $3.6\times10^{-3}$ & $9.0\times10^{-3}$ & $1.1\times10^{-2}$ & $8.1\times10^{-3}$ & $2.1\times10^{-2}$ & $4.2\times10^{-3}$ & $7.0\times10^{-2}$ \\
        $^{15}$N & $4.1\times10^{-5}$ & $6.2\times10^{-4}$ & $7.0\times10^{-4}$ & $1.8\times10^{-4}$ & $2.1\times10^{-3}$ & $1.1\times10^{-4}$ & $3.9\times10^{-4}$ & $1.5\times10^{-5}$ & $1.2\times10^{-6}$ & $1.3\times10^{-3}$ \\
        $^{16}$O & $1.1\times10^{-1}$ & $2.4\times10^{-1}$ & $1.50$ & $7.1\times10^{-1}$ & $2.15$ & $1.52$ & $1.10$ & $1.39$ & $4.4\times10^{-1}$ & $8.0\times10^{-1}$ \\
        $^{17}$O & $1.9\times10^{-5}$ & $1.7\times10^{-5}$ & $2.9\times10^{-5}$ & $1.1\times10^{-5}$ & $7.6\times10^{-4}$ & $5.5\times10^{-5}$ & $3.3\times10^{-5}$ & $5.8\times10^{-4}$ & $5.8\times10^{-7}$ & $2.5\times10^{-5}$ \\
        $^{18}$O & $3.0\times10^{-5}$ & $2.6\times10^{-5}$ & $4.6\times10^{-3}$ & $8.3\times10^{-6}$ & $4.2\times10^{-6}$ & $3.9\times10^{-7}$ & $4.7\times10^{-7}$ & $9.7\times10^{-8}$ & $4.2\times10^{-8}$ & $5.0\times10^{-8}$ \\
        $^{18}$F & $0$ & $0$ & $0$ & $1.1\times10^{-8}$ & $7.8\times10^{-6}$ & $3.0\times10^{-9}$ & $3.2\times10^{-8}$ & $8.3\times10^{-8}$ & $0$ & $4.9\times10^{-9}$ \\
        $^{19}$F & $1.3\times10^{-6}$ & $1.3\times10^{-6}$ & $5.2\times10^{-5}$ & $1.2\times10^{-7}$ & $3.4\times10^{-6}$ & $1.8\times10^{-6}$ & $1.1\times10^{-6}$ & $1.2\times10^{-6}$ & $5.3\times10^{-6}$ & $1.9\times10^{-8}$ \\
        $^{20}$Ne & $5.9\times10^{-5}$ & $1.1\times10^{-3}$ & $3.0\times10^{-2}$ & $1.6\times10^{-1}$ & $1.4\times10^{-1}$ & $2.2\times10^{-1}$ & $2.0\times10^{-2}$ & $1.4\times10^{-2}$ & $1.6\times10^{-3}$ & $1.5\times10^{-1}$ \\
        $^{21}$Ne & $7.4\times10^{-7}$ & $4.7\times10^{-6}$ & $3.0\times10^{-4}$ & $8.3\times10^{-7}$ & $2.5\times10^{-4}$ & $3.0\times10^{-5}$ & $1.4\times10^{-5}$ & $1.9\times10^{-5}$ & $2.7\times10^{-5}$ & $8.5\times10^{-6}$ \\
        $^{22}$Ne & $1.9\times10^{-4}$ & $1.9\times10^{-4}$ & $2.1\times10^{-2}$ & $2.3\times10^{-5}$ & $2.2\times10^{-4}$ & $3.4\times10^{-3}$ & $6.5\times10^{-4}$ & $1.0\times10^{-3}$ & $5.4\times10^{-3}$ & $3.5\times10^{-5}$ \\
        $^{23}$Na & $1.3\times10^{-6}$ & $1.5\times10^{-6}$ & $1.8\times10^{-4}$ & $5.1\times10^{-7}$ & $3.7\times10^{-6}$ & $3.5\times10^{-5}$ & $1.1\times10^{-5}$ & $2.2\times10^{-5}$ & $4.2\times10^{-5}$ & $2.9\times10^{-7}$ \\
        $^{24}$Mg & $1.5\times10^{-6}$ & $5.7\times10^{-5}$ & $3.0\times10^{-3}$ & $9.0\times10^{-3}$ & $1.5\times10^{-2}$ & $1.5\times10^{-2}$ & $8.5\times10^{-4}$ & $5.6\times10^{-4}$ & $2.1\times10^{-5}$ & $3.2\times10^{-2}$ \\
        $^{25}$Mg & $1.8\times10^{-6}$ & $1.4\times10^{-5}$ & $2.2\times10^{-3}$ & $6.9\times10^{-7}$ & $1.2\times10^{-3}$ & $6.3\times10^{-5}$ & $1.3\times10^{-5}$ & $1.9\times10^{-5}$ & $8.7\times10^{-5}$ & $5.0\times10^{-5}$ \\
        $^{26}$Mg & $4.8\times10^{-6}$ & $2.2\times10^{-5}$ & $5.4\times10^{-3}$ & $1.2\times10^{-6}$ & $4.6\times10^{-4}$ & $1.4\times10^{-4}$ & $1.6\times10^{-5}$ & $3.2\times10^{-4}$ & $1.3\times10^{-4}$ & $2.0\times10^{-5}$ \\
        $^{26}$Al & $0$ & $0$ & $0$ & $0$ & $0$ & $7.9\times10^{-10}$ & $0$ & $6.2\times10^{-10}$ & $1.4\times10^{-8}$ & $3.5\times10^{-9}$ \\
        $^{27}$Al & $3.0\times10^{-8}$ & $9.5\times10^{-8}$ & $3.9\times10^{-5}$ & $7.2\times10^{-9}$ & $1.7\times10^{-6}$ & $8.8\times10^{-7}$ & $6.8\times10^{-8}$ & $5.7\times10^{-6}$ & $5.1\times10^{-7}$ & $1.0\times10^{-8}$ \\
        $^{28}$Si & $0$ & $0$ & $0$ & $1.8\times10^{-4}$ & $7.9\times10^{-4}$ & $3.7\times10^{-4}$ & $1.1\times10^{-5}$ & $0$ & $0$ & $1.2\times10^{-6}$ \\
        $^{29}$Si & $0$ & $0$ & $0$ & $0$ & $0$ & $0$ & $0$ & $0$ & $0$ & $0$ \\
        $^{30}$Si & $0$ & $0$ & $0$ & $0$ & $0$ & $0$ & $0$ & $0$ & $0$ & $0$ \\
        $^{31}$P & $0$ & $0$ & $0$ & $0$ & $0$ & $0$ & $0$ & $0$ & $0$ & $0$ \\
        $^{32}$S & $0$ & $0$ & $0$ & $1.8\times10^{-7}$ & $2.4\times10^{-5}$ & $7.0\times10^{-7}$ & $1.6\times10^{-8}$ & $0$ & $0$ & $2.9\times10^{-9}$ \\
        $^{33}$S & $0$ & $0$ & $0$ & $0$ & $0$ & $0$ & $0$ & $0$ & $0$ & $0$ \\
        $^{34}$S & $0$ & $0$ & $0$ & $0$ & $0$ & $0$ & $0$ & $0$ & $0$ & $0$ \\
        $^{36}$S & $0$ & $0$ & $0$ & $0$ & $0$ & $0$ & $0$ & $0$ & $0$ & $0$ \\
        $^{35}$Cl & $0$ & $0$ & $0$ & $0$ & $0$ & $0$ & $0$ & $0$ & $0$ & $0$ \\
        $^{37}$Cl & $0$ & $0$ & $0$ & $0$ & $0$ & $0$ & $0$ & $0$ & $0$ & $0$ \\
        $^{36}$Ar & $0$ & $0$ & $0$ & $0$ & $7.1\times10^{-8}$ & $0$ & $0$ & $0$ & $0$ & $0$ \\
        $^{38}$Ar & $0$ & $0$ & $0$ & $0$ & $0$ & $0$ & $0$ & $0$ & $0$ & $0$ \\
        $^{39}$K & $0$ & $0$ & $0$ & $0$ & $0$ & $0$ & $0$ & $0$ & $0$ & $0$ \\
        $^{40}$K & $0$ & $0$ & $0$ & $0$ & $0$ & $0$ & $0$ & $0$ & $0$ & $0$ \\
        $^{41}$K & $0$ & $0$ & $0$ & $0$ & $0$ & $0$ & $0$ & $0$ & $0$ & $0$ \\
        $^{40}$Ca & $0$ & $0$ & $0$ & $0$ & $0$ & $0$ & $0$ & $0$ & $0$ & $0$ \\
        $^{42}$Ca & $0$ & $0$ & $0$ & $0$ & $0$ & $0$ & $0$ & $0$ & $0$ & $0$ \\
        $^{43}$Ca & $0$ & $0$ & $0$ & $0$ & $0$ & $0$ & $0$ & $0$ & $0$ & $0$ \\
        $^{44}$Ca & $0$ & $0$ & $0$ & $0$ & $0$ & $0$ & $0$ & $0$ & $0$ & $0$ \\
        $^{45}$Sc & $0$ & $0$ & $0$ & $0$ & $0$ & $0$ & $0$ & $0$ & $0$ & $0$ \\
        $^{46}$Ti & $0$ & $0$ & $0$ & $0$ & $0$ & $0$ & $0$ & $0$ & $0$ & $0$ \\
        $^{47}$Ti & $0$ & $0$ & $0$ & $0$ & $0$ & $0$ & $0$ & $0$ & $0$ & $0$ \\
        $^{48}$Ti & $0$ & $0$ & $0$ & $0$ & $0$ & $0$ & $0$ & $0$ & $0$ & $0$ \\
        $^{49}$Ti & $0$ & $0$ & $0$ & $0$ & $0$ & $0$ & $0$ & $0$ & $0$ & $0$ \\
        $^{51}$V & $0$ & $0$ & $0$ & $0$ & $0$ & $0$ & $0$ & $0$ & $0$ & $0$ \\
        $^{50}$Cr & $0$ & $0$ & $0$ & $0$ & $0$ & $0$ & $0$ & $0$ & $0$ & $0$ \\
        $^{52}$Cr & $0$ & $0$ & $0$ & $0$ & $0$ & $0$ & $0$ & $0$ & $0$ & $0$ \\
        $^{53}$Cr & $0$ & $0$ & $0$ & $0$ & $0$ & $0$ & $0$ & $0$ & $0$ & $0$ \\
        $^{54}$Cr & $0$ & $0$ & $0$ & $0$ & $0$ & $0$ & $0$ & $0$ & $0$ & $0$ \\
        $^{55}$Mn & $0$ & $0$ & $0$ & $0$ & $0$ & $0$ & $0$ & $0$ & $0$ & $0$ \\
        $^{54}$Fe & $0$ & $0$ & $0$ & $0$ & $0$ & $0$ & $0$ & $0$ & $0$ & $0$ \\
        $^{56}$Fe & $0$ & $0$ & $0$ & $0$ & $0$ & $0$ & $0$ & $0$ & $0$ & $0$ \\
        $^{57}$Fe & $0$ & $0$ & $0$ & $0$ & $0$ & $0$ & $0$ & $0$ & $0$ & $0$ \\
        $^{58}$Fe & $0$ & $0$ & $0$ & $0$ & $0$ & $0$ & $0$ & $0$ & $0$ & $0$ \\
        $^{59}$Co & $0$ & $0$ & $0$ & $0$ & $0$ & $0$ & $0$ & $0$ & $0$ & $0$ \\
        $^{58}$Ni & $0$ & $0$ & $0$ & $0$ & $0$ & $0$ & $0$ & $0$ & $0$ & $0$ \\
        $^{60}$Ni & $0$ & $0$ & $0$ & $0$ & $0$ & $0$ & $0$ & $0$ & $0$ & $0$ \\
        $^{61}$Ni & $0$ & $0$ & $0$ & $0$ & $0$ & $0$ & $0$ & $0$ & $0$ & $0$ \\
        $^{62}$Ni & $0$ & $0$ & $0$ & $0$ & $0$ & $0$ & $0$ & $0$ & $0$ & $0$ \\
        $^{63}$Cu & $0$ & $0$ & $0$ & $0$ & $0$ & $0$ & $0$ & $0$ & $0$ & $0$ \\
        $^{64}$Zn & $0$ & $0$ & $0$ & $0$ & $0$ & $0$ & $0$ & $0$ & $0$ & $0$ \\
        $^{66}$Zn & $0$ & $0$ & $0$ & $0$ & $0$ & $0$ & $0$ & $0$ & $0$ & $0$ \\
        \hline
    \end{tabular}
    }
\end{table*}

\begin{table*}
    \caption{Gross adopted ejecta masses for selected isotopes for selected isotopes as a function of initial stellar mass ($M_\odot$) over the mass range $60$--$200\,M_\odot$. Yields below $10^{-10}\,M_\odot$ are treated as zero. For CC and PPISN models, species heavier than Ne are stage dependent reservoirs and not final explosive nucleosynthesis yields.}
    \label{tab:yields_highmass}
    \scriptsize                 
    \setlength{\tabcolsep}{6pt}    
    \renewcommand{\arraystretch}{1}      
    \resizebox{\textwidth}{!}{
    \hspace*{-0.17\textwidth}
    \begin{tabular}{|l|cccccccccc|}
        \hline
        Isotope & \multicolumn{10}{c|}{Stellar Mass ($M_\odot$)} \\
        \cline{2-11} & $60$ & $65$ & $70$ & $80$ & $90$ & $100$ & $120$ & $150$ & $180$ & $200$ \\
        \hline
        $^{1}$H & $12.7$ & $13.9$ & $14.8$ & $15.9$ & $15.8$ & $18.2$ & $19.3$ & $20.8$ & $21.5$ & $22.9$ \\
        $^{3}$He & $8.5\times10^{-3}$ & $1.5\times10^{-3}$ & $9.0\times10^{-3}$ & $6.7\times10^{-3}$ & $7.3\times10^{-4}$ & $0$ & $0$ & $0$ & $0$ & $0$ \\
        $^{4}$He & $16.4$ & $21.2$ & $21.6$ & $25.4$ & $25.4$ & $28.7$ & $38.0$ & $49.6$ & $57.6$ & $55.8$ \\
        $^{12}$C & $2.4\times10^{-2}$ & $4.3\times10^{-1}$ & $1.08$ & $6.5\times10^{-1}$ & $7.3\times10^{-1}$ & $8.6\times10^{-1}$ & $4.21$ & $11.3$ & $9.00$ & $6.89$ \\
        $^{13}$C & $3.0\times10^{-5}$ & $1.2\times10^{-2}$ & $1.2\times10^{-1}$ & $8.0\times10^{-2}$ & $9.2\times10^{-2}$ & $1.7\times10^{-3}$ & $4.8\times10^{-1}$ & $6.5\times10^{-1}$ & $8.0\times10^{-1}$ & $6.7\times10^{-1}$ \\
        $^{14}$C & $5.5\times10^{-3}$ & $3.0\times10^{-5}$ & $4.2\times10^{-7}$ & $2.3\times10^{-5}$ & $2.2\times10^{-6}$ & $0$ & $0$ & $0$ & $0$ & $0$ \\
        $^{14}$N & $1.7\times10^{-3}$ & $1.2\times10^{-1}$ & $2.7\times10^{-1}$ & $4.0\times10^{-1}$ & $3.4\times10^{-1}$ & $8.9\times10^{-3}$ & $1.23$ & $1.81$ & $2.52$ & $2.42$ \\
        $^{15}$N & $1.3\times10^{-4}$ & $9.1\times10^{-4}$ & $1.4\times10^{-2}$ & $1.3\times10^{-2}$ & $5.6\times10^{-3}$ & $0$ & $0$ & $7.3\times10^{-7}$ & $6.6\times10^{-6}$ & $6.4\times10^{-6}$ \\
        $^{16}$O & $3.3\times10^{-2}$ & $4.3\times10^{-1}$ & $2.23$ & $3.32$ & $5.01$ & $5.89$ & $9.50$ & $57.0$ & $55.6$ & $65.4$ \\
        $^{17}$O & $2.4\times10^{-7}$ & $5.7\times10^{-5}$ & $3.1\times10^{-4}$ & $1.2\times10^{-3}$ & $1.3\times10^{-3}$ & $1.4\times10^{-5}$ & $1.4\times10^{-3}$ & $2.1\times10^{-3}$ & $3.0\times10^{-3}$ & $1.1\times10^{-2}$ \\
        $^{18}$O & $9.5\times10^{-8}$ & $5.6\times10^{-6}$ & $3.8\times10^{-7}$ & $3.1\times10^{-6}$ & $1.8\times10^{-6}$ & $2.6\times10^{-5}$ & $1.7\times10^{-6}$ & $9.8\times10^{-6}$ & $1.4\times10^{-5}$ & $1.7\times10^{-5}$ \\
        $^{18}$F & $3.3\times10^{-7}$ & $1.5\times10^{-5}$ & $1.3\times10^{-6}$ & $7.7\times10^{-9}$ & $0$ & $0$ & $0$ & $0$ & $0$ & $0$ \\
        $^{19}$F & $1.1\times10^{-6}$ & $3.2\times10^{-8}$ & $1.2\times10^{-8}$ & $1.5\times10^{-5}$ & $4.5\times10^{-8}$ & $4.3\times10^{-9}$ & $5.7\times10^{-8}$ & $1.0\times10^{-4}$ & $1.1\times10^{-5}$ & $5.7\times10^{-7}$ \\
        $^{20}$Ne & $6.2\times10^{-3}$ & $4.0\times10^{-4}$ & $7.4\times10^{-3}$ & $7.1\times10^{-2}$ & $1.4\times10^{-1}$ & $5.8\times10^{-1}$ & $1.4\times10^{-1}$ & $5.01$ & $4.22$ & $4.26$ \\
        $^{21}$Ne & $7.1\times10^{-6}$ & $6.9\times10^{-7}$ & $4.9\times10^{-6}$ & $1.4\times10^{-5}$ & $2.9\times10^{-5}$ & $3.7\times10^{-5}$ & $2.4\times10^{-5}$ & $7.4\times10^{-4}$ & $5.7\times10^{-4}$ & $5.5\times10^{-4}$ \\
        $^{22}$Ne & $5.8\times10^{-5}$ & $1.0\times10^{-5}$ & $3.4\times10^{-5}$ & $2.6\times10^{-5}$ & $3.7\times10^{-5}$ & $4.0\times10^{-5}$ & $1.3\times10^{-4}$ & $1.1\times10^{-3}$ & $1.5\times10^{-3}$ & $7.4\times10^{-4}$ \\
        $^{23}$Na & $7.8\times10^{-8}$ & $2.0\times10^{-7}$ & $8.9\times10^{-7}$ & $9.7\times10^{-7}$ & $1.8\times10^{-6}$ & $1.8\times10^{-6}$ & $3.6\times10^{-6}$ & $9.1\times10^{-3}$ & $3.0\times10^{-3}$ & $2.8\times10^{-3}$ \\
        $^{24}$Mg & $1.3\times10^{-3}$ & $3.7\times10^{-6}$ & $1.2\times10^{-4}$ & $5.6\times10^{-3}$ & $1.2\times10^{-2}$ & $1.4\times10^{-1}$ & $1.4\times10^{-2}$ & $1.53$ & $3.79$ & $4.40$ \\
        $^{25}$Mg & $6.3\times10^{-5}$ & $6.6\times10^{-7}$ & $8.5\times10^{-6}$ & $3.2\times10^{-5}$ & $8.1\times10^{-5}$ & $1.5\times10^{-4}$ & $4.4\times10^{-5}$ & $3.8\times10^{-3}$ & $2.7\times10^{-3}$ & $5.5\times10^{-3}$ \\
        $^{26}$Mg & $2.8\times10^{-5}$ & $8.0\times10^{-7}$ & $9.6\times10^{-6}$ & $2.7\times10^{-5}$ & $6.2\times10^{-5}$ & $9.0\times10^{-5}$ & $4.8\times10^{-5}$ & $3.8\times10^{-3}$ & $1.9\times10^{-3}$ & $2.9\times10^{-3}$ \\
        $^{26}$Al & $9.1\times10^{-9}$ & $3.5\times10^{-9}$ & $4.6\times10^{-9}$ & $1.7\times10^{-8}$ & $2.2\times10^{-8}$ & $7.1\times10^{-9}$ & $8.3\times10^{-9}$ & $0$ & $0$ & $0$ \\
        $^{27}$Al & $3.8\times10^{-8}$ & $1.2\times10^{-8}$ & $3.0\times10^{-8}$ & $6.0\times10^{-8}$ & $9.2\times10^{-8}$ & $8.9\times10^{-8}$ & $1.0\times10^{-7}$ & $3.4\times10^{-2}$ & $1.6\times10^{-2}$ & $1.7\times10^{-2}$ \\
        $^{28}$Si & $0$ & $0$ & $0$ & $0$ & $0$ & $5.9\times10^{-10}$ & $1.4\times10^{-4}$ & $3.2\times10^{-1}$ & $14.8$ & $21.7$ \\
        $^{29}$Si & $0$ & $0$ & $0$ & $0$ & $0$ & $0$ & $0$ & $2.0\times10^{-3}$ & $2.2\times10^{-2}$ & $2.0\times10^{-2}$ \\
        $^{30}$Si & $0$ & $0$ & $0$ & $0$ & $0$ & $0$ & $0$ & $6.5\times10^{-3}$ & $1.3\times10^{-3}$ & $1.3\times10^{-3}$ \\
        $^{31}$P & $0$ & $0$ & $0$ & $0$ & $0$ & $0$ & $0$ & $2.7\times10^{-3}$ & $1.8\times10^{-3}$ & $1.3\times10^{-3}$ \\
        $^{32}$S & $0$ & $0$ & $0$ & $0$ & $0$ & $0$ & $8.1\times10^{-7}$ & $9.5\times10^{-4}$ & $5.27$ & $9.06$ \\
        $^{33}$S & $0$ & $0$ & $0$ & $0$ & $0$ & $0$ & $0$ & $3.8\times10^{-5}$ & $3.7\times10^{-3}$ & $3.7\times10^{-3}$ \\
        $^{34}$S & $0$ & $0$ & $0$ & $0$ & $0$ & $0$ & $0$ & $2.1\times10^{-4}$ & $2.5\times10^{-3}$ & $7.3\times10^{-4}$ \\
        $^{36}$S & $0$ & $0$ & $0$ & $0$ & $0$ & $0$ & $0$ & $1.1\times10^{-8}$ & $6.2\times10^{-9}$ & $3.7\times10^{-9}$ \\
        $^{35}$Cl & $0$ & $0$ & $0$ & $0$ & $0$ & $0$ & $0$ & $5.6\times10^{-5}$ & $7.0\times10^{-4}$ & $3.8\times10^{-4}$ \\
        $^{37}$Cl & $0$ & $0$ & $0$ & $0$ & $0$ & $0$ & $0$ & $1.1\times10^{-7}$ & $6.8\times10^{-4}$ & $6.8\times10^{-4}$ \\
        $^{36}$Ar & $0$ & $0$ & $0$ & $0$ & $0$ & $0$ & $3.8\times10^{-10}$ & $1.3\times10^{-5}$ & $6.8\times10^{-1}$ & $1.36$ \\
        $^{38}$Ar & $0$ & $0$ & $0$ & $0$ & $0$ & $0$ & $0$ & $4.9\times10^{-7}$ & $2.9\times10^{-3}$ & $8.3\times10^{-4}$ \\
        $^{39}$K & $0$ & $0$ & $0$ & $0$ & $0$ & $0$ & $0$ & $1.3\times10^{-7}$ & $9.0\times10^{-4}$ & $4.8\times10^{-4}$ \\
        $^{40}$K & $0$ & $0$ & $0$ & $0$ & $0$ & $0$ & $0$ & $2.5\times10^{-10}$ & $2.0\times10^{-8}$ & $9.2\times10^{-9}$ \\
        $^{41}$K & $0$ & $0$ & $0$ & $0$ & $0$ & $0$ & $0$ & $2.4\times10^{-10}$ & $1.5\times10^{-4}$ & $1.5\times10^{-4}$ \\
        $^{40}$Ca & $0$ & $0$ & $0$ & $0$ & $0$ & $0$ & $0$ & $1.5\times10^{-8}$ & $4.4\times10^{-1}$ & $1.04$ \\
        $^{42}$Ca & $0$ & $0$ & $0$ & $0$ & $0$ & $0$ & $0$ & $2.4\times10^{-10}$ & $7.5\times10^{-5}$ & $2.2\times10^{-5}$ \\
        $^{43}$Ca & $0$ & $0$ & $0$ & $0$ & $0$ & $0$ & $0$ & $0$ & $2.4\times10^{-8}$ & $9.5\times10^{-9}$ \\
        $^{44}$Ca & $0$ & $0$ & $0$ & $0$ & $0$ & $0$ & $0$ & $0$ & $9.6\times10^{-5}$ & $3.0\times10^{-4}$ \\
        $^{45}$Sc & $0$ & $0$ & $0$ & $0$ & $0$ & $0$ & $0$ & $0$ & $3.1\times10^{-6}$ & $3.5\times10^{-6}$ \\
        $^{46}$Ti & $0$ & $0$ & $0$ & $0$ & $0$ & $0$ & $0$ & $0$ & $4.1\times10^{-5}$ & $1.8\times10^{-5}$ \\
        $^{47}$Ti & $0$ & $0$ & $0$ & $0$ & $0$ & $0$ & $0$ & $0$ & $3.6\times10^{-7}$ & $6.8\times10^{-7}$ \\
        $^{48}$Ti & $0$ & $0$ & $0$ & $0$ & $0$ & $0$ & $0$ & $0$ & $9.4\times10^{-4}$ & $6.6\times10^{-3}$ \\
        $^{49}$Ti & $0$ & $0$ & $0$ & $0$ & $0$ & $0$ & $0$ & $0$ & $4.6\times10^{-5}$ & $1.9\times10^{-4}$ \\
        $^{51}$V & $0$ & $0$ & $0$ & $0$ & $0$ & $0$ & $0$ & $0$ & $8.4\times10^{-5}$ & $3.2\times10^{-4}$ \\
        $^{50}$Cr & $0$ & $0$ & $0$ & $0$ & $0$ & $0$ & $0$ & $0$ & $3.9\times10^{-4}$ & $5.0\times10^{-4}$ \\
        $^{52}$Cr & $0$ & $0$ & $0$ & $0$ & $0$ & $0$ & $0$ & $0$ & $1.4\times10^{-2}$ & $1.4\times10^{-1}$ \\
        $^{53}$Cr & $0$ & $0$ & $0$ & $0$ & $0$ & $0$ & $0$ & $0$ & $9.2\times10^{-4}$ & $5.7\times10^{-3}$ \\
        $^{54}$Cr & $0$ & $0$ & $0$ & $0$ & $0$ & $0$ & $0$ & $0$ & $9.7\times10^{-9}$ & $4.3\times10^{-9}$ \\
        $^{55}$Mn & $0$ & $0$ & $0$ & $0$ & $0$ & $0$ & $0$ & $0$ & $3.9\times10^{-3}$ & $2.3\times10^{-2}$ \\
        $^{54}$Fe & $0$ & $0$ & $0$ & $0$ & $0$ & $0$ & $0$ & $0$ & $4.4\times10^{-2}$ & $9.6\times10^{-2}$ \\
        $^{56}$Fe & $0$ & $0$ & $0$ & $0$ & $0$ & $0$ & $0$ & $0$ & $2.4\times10^{-1}$ & $3.51$ \\
        $^{57}$Fe & $0$ & $0$ & $0$ & $0$ & $0$ & $0$ & $0$ & $0$ & $1.8\times10^{-3}$ & $1.6\times10^{-2}$ \\
        $^{58}$Fe & $0$ & $0$ & $0$ & $0$ & $0$ & $0$ & $0$ & $0$ & $8.8\times10^{-9}$ & $6.4\times10^{-9}$ \\
        $^{59}$Co & $0$ & $0$ & $0$ & $0$ & $0$ & $0$ & $0$ & $0$ & $4.4\times10^{-6}$ & $7.0\times10^{-6}$ \\
        $^{58}$Ni & $0$ & $0$ & $0$ & $0$ & $0$ & $0$ & $0$ & $0$ & $2.8\times10^{-3}$ & $1.0\times10^{-2}$ \\
        $^{60}$Ni & $0$ & $0$ & $0$ & $0$ & $0$ & $0$ & $0$ & $0$ & $6.5\times10^{-6}$ & $1.1\times10^{-5}$ \\
        $^{61}$Ni & $0$ & $0$ & $0$ & $0$ & $0$ & $0$ & $0$ & $0$ & $9.2\times10^{-10}$ & $3.6\times10^{-7}$ \\
        $^{62}$Ni & $0$ & $0$ & $0$ & $0$ & $0$ & $0$ & $0$ & $0$ & $2.2\times10^{-10}$ & $1.5\times10^{-6}$ \\
        $^{63}$Cu & $0$ & $0$ & $0$ & $0$ & $0$ & $0$ & $0$ & $0$ & $0$ & $5.5\times10^{-10}$ \\
        $^{64}$Zn & $0$ & $0$ & $0$ & $0$ & $0$ & $0$ & $0$ & $0$ & $0$ & $8.1\times10^{-10}$ \\
        $^{66}$Zn & $0$ & $0$ & $0$ & $0$ & $0$ & $0$ & $0$ & $0$ & $0$ & $1.0\times10^{-9}$ \\ 
        \hline
    \end{tabular}
    }
\end{table*}

The closest direct MESA comparison is the $25,M_\odot$ Pop~III study of \citet{Aryan2023}, who evolved models with $\Omega/\Omega_{\rm crit}$ from 0 to 0.8 to the onset of iron core collapse. Their models with $\Omega/\Omega_{\rm crit}=0.6$ and 0.8 experienced strong mass loss after the main sequence and reached final masses of $11.94$ and $11.79,M_\odot$, respectively. These models also became compact and strongly stripped, with final radii of $1.5$ and $0.6,R_\odot$. Our $25,M_\odot$ model with $\upsilon_{\rm ini}/\upsilon_{\rm crit}=0.7$ instead retains an H and He rich envelope, while its surface CNO mass fraction remains below a few $10^{-9}$. Its adopted ejecta contain $8.11,M_\odot$ of H, $10.3,M_\odot$ of He, $5.6\times10^{-1},M_\odot$ of $^{12}$C, $8.1\times10^{-3},M_\odot$ of $^{14}$N, and $1.10,M_\odot$ of $^{16}$O. Both calculations show that rapid rotation can strongly influence the final structure, but they differ substantially in the amount of stripping. This difference reflects the combined effects of the rotational transport and mass loss prescriptions. The MESA models use the Dutch wind prescription with a scaling factor of 0.5 and allow surface CNO enrichment to activate strong radiative mass loss, while our models exclude line driven winds at $Z=0$ and lose mass only through the mechanical channel near critical rotation. A controlled calculation with otherwise identical input physics would be required to isolate the contribution of each assumption. The subsequent SNEC calculations of \citet{Aryan2023} do not provide an elemental yield comparison because the central remnant is excised, the explosion energy and nickel mass are prescribed, and SNEC does not include a nuclear reaction network.

A broader MESA comparison is provided by \citet{Hassan2025}, who calculated 64 Pop~III models with initial masses from $10$ to $800,M_\odot$ and rotation rates of 0, 0.2, 0.4, and 0.6 of the critical velocity. Their calculations show that rotation has little effect on observability at the ZAMS, while subsequent rotational evolution can maintain hotter and more luminous stellar configurations. Chemically homogeneous evolution occurs in part of their rapidly rotating grid, although rotational mass loss prevents this behaviour in some of the most massive models. Our calculations similarly show that rotation changes the evolutionary tracks and internal chemical transport, but the mass dependence is different. The $5$ and $7,M_\odot$ models in our grid come closest to chemically homogeneous evolution, the models from $9$ to $20,M_\odot$ show only a temporary tendency toward this behaviour, and the models above $20,M_\odot$ evolve redward during core H burning. The comparison is not one to one because the MESA models begin on the pre main sequence, adopt a maximum initial rotation of 0.6 of critical, and were primarily analysed through core H exhaustion for their spectral properties. Our models begin at the ZAMS with an initial rotation of 0.7 of critical and are followed through the later burning stages. The differences are therefore consistent with the different initial conditions, rotational transport prescriptions, mixing calibrations, and mass loss treatments adopted in the two calculations.

\section{Conclusions}
\label{sec:Conclusions}

We have presented a new CNO-focused ejecta dataset for rapidly rotating Pop~III stars with $5 \leq M_{\rm ini}/M_\odot \leq 200$ and $\upsilon_{\rm ini}/\upsilon_{\rm crit}=0.7$. The models were evolved from the ZAMS through core O burning where applicable, and their calculated abundance profiles were converted into gross adopted ejecta masses using fate-dependent remnant boundaries. The principal nucleosynthetic product of this work is the hydrostatically produced CNO material tabulated across the CC, PPISN, and PISN regimes. Species heavier than Ne in the CC and PPISN tables are provided as stage dependent final model reservoirs rather than final hydrodynamic explosion yields.

Because the vast majority of stars between $30$ and $90\,M_{\odot}$ are expected to collapse directly to black holes without a successful explosion, we also provide a second yield set in Tables~\ref{tab:isotope_yields_windonly:p1} and~\ref{tab:isotope_yields_windonly:p2} that adopts this assumption explicitly. Over $30$--$90\,M_{\odot}$, these tables include only the pre-collapse stellar-wind ejecta, with all interior processed layers assumed to be swallowed by the remnant rather than ejected. Outside this mass range, Tables~\ref{tab:isotope_yields_windonly:p1} and~\ref{tab:isotope_yields_windonly:p2} are identical to Tables~\ref{tab:yields_lowmass} and~\ref{tab:yields_highmass}. This second set isolates how much of the CNO and $\alpha$-element budget in our fiducial tables depends on the working assumption that these intermediate-mass core-collapse models explode at all.

The grid shows that fast rotation can seed the ejecta with primary CNO material over a broad mass range, while the final fate determines how much deeper processed material accompanies it. In the core-collapse and PPISN branches, the adopted ejecta are mainly H/He-rich material and outer CNO-enriched layers. The compact O/Ne/Mg and Si/S/Ar/Ca-rich regions are mostly retained. In the PISN branch, the full processed core is released. The yields therefore change from CNO-rich retained-core enrichment to O-rich and deep-alpha-rich full-disruption enrichment.

This distinction is essential for nitrogen. The grid confirms that fast rotation produces primary N at zero metallicity, but the yield tables show that nitrogen mass and N/O are not the same diagnostic. The PISN models eject the largest absolute N masses, yet they also release tens of solar masses of O. Their ejecta are therefore not the most N/O-enhanced. The highest N/O values instead occur in selected retained-core models, where less N is ejected in absolute mass but a large fraction of the O-rich core remains locked in the remnant. Thus high N production does not automatically imply high N/O.

The IMF-weighted mixtures give the same result at population level. A more top-heavy weighting increases the contribution from PISN ejecta and therefore adds O and deep-alpha material faster than it increases N/O. The integrated abundance pattern moves toward a PISN-like mixture rather than a N-dominated one. This provides a useful constraint on early chemical enrichment models: a top-heavy population of $5$--$200\,M_\odot$ fast rotators is not, by itself, a solution to the most extreme high-redshift N abundances.

Using the assumed remnant masses prescriptions, the present yields can contribute to CNO-rich enrichment in the early Universe, but they do not reproduce the most N-dominated JWST systems on their own. The maximum N/O ratio in the grid remains below the values inferred for GN-z11, CEERS-1019, and GS~3073 \citep{Cameron2023,Marques2023,Ji2024,Ji2026}. These sources likely require selective enrichment by the strongest retained-core N producers, or an additional very massive to supermassive Pop~III channel \citep{Nandal2025}. The value of the present grid is therefore to define the CNO enrichment that ordinary massive fast rotators can supply, and to identify where a different enrichment channel is required.

The next step is to replace the parametric ejecta--remnant partition with self-consistent collapse and explosion calculations. This is needed to follow fallback, shock processing, explosive burning, and Fe-group production in the core-collapse branch. The PPISN models should also be followed through pulse hydrodynamics rather than through a net remnant mass prescription \citep{Heger2002,Woosley2017}. Future grids should test how internal magnetic angular-momentum transport changes the final core structure, for example through the Tayler--Spruit dynamo \citep{Tayler1973,Spruit2002}. Extending the same yield analysis to very massive and supermassive Pop~III stars will be essential for testing the origin of the highest N/O ratios now observed at high redshift.

\begin{acknowledgments}
DN was supported by the Swiss National Science Fund (SNSF) Postdoctoral Fellowship, grant number: P500-2235464.  
RH acknowledges support from STFC, the World Premier International Research Centre Initiative (WPI Initiative), MEXT, Japan, the IReNA AccelNet Network of Networks (NSF, Grant No. OISE-1927130), CeNAM (grant DE-SC0026204) and {the Wolfson Foundation that part-funded the greenHPC facility at Keele}.
\end{acknowledgments}

\bibliography{sample7}{}
\bibliographystyle{aasjournalv7}

\begin{table*}[htbp]
    \caption{Gross adopted ejecta masses for selected isotopes} for individual isotopes as a function of initial stellar mass ($M_\odot$). For $30 \leq M/M_\odot \leq 50$, only pre-SN stellar wind ejecta are included (terminal SN ejecta for this range are assumed to fall into the remnant); outside this range, total ejecta = terminal SN/PPISN/PISN + pre-SN stellar winds. Entries marked `$0$' fall below $10^{-10}\,M_\odot$.
    \label{tab:isotope_yields_windonly:p1}
    \scriptsize                 
    \setlength{\tabcolsep}{6pt}    
    \renewcommand{\arraystretch}{1}      
    \resizebox{\textwidth}{!}{
    \hspace*{-0.17\textwidth}
    \begin{tabular}{|l|cccccccccc|}
        \hline
        Isotope & \multicolumn{10}{c|}{Stellar Mass ($M_\odot$)} \\
        \cline{2-11} & $5$ & $7$ & $9$ & $12$ & $15$ & $20$ & $25$ & $30$ & $40$ & $50$ \\
        \hline
        $^{1}$H & $1.33$ & $2.33$ & $3.33$ & $4.69$ & $4.90$ & $6.25$ & $8.11$ & $2.3\times10^{-1}$ & $4.2\times10^{-1}$ & $7.6\times10^{-1}$ \\
        $^{3}$He & $3.0\times10^{-5}$ & $6.4\times10^{-4}$ & $2.9\times10^{-4}$ & $1.9\times10^{-3}$ & $5.1\times10^{-3}$ & $1.1\times10^{-3}$ & $1.7\times10^{-3}$ & $0$ & $0$ & $0$ \\
        $^{4}$He & $2.01$ & $2.92$ & $2.32$ & $4.83$ & $6.15$ & $9.14$ & $10.3$ & $8.8\times10^{-2}$ & $1.5\times10^{-1}$ & $2.7\times10^{-1}$ \\
        $^{12}$C & $7.9\times10^{-2}$ & $1.3\times10^{-1}$ & $2.5\times10^{-1}$ & $2.6\times10^{-1}$ & $1.9\times10^{-1}$ & $6.0\times10^{-1}$ & $5.6\times10^{-1}$ & $0$ & $0$ & $0$ \\
        $^{13}$C & $1.1\times10^{-3}$ & $1.7\times10^{-3}$ & $1.5\times10^{-3}$ & $4.2\times10^{-3}$ & $1.2\times10^{-2}$ & $2.0\times10^{-2}$ & $7.0\times10^{-3}$ & $0$ & $0$ & $0$ \\
        $^{14}$C & $3.6\times10^{-6}$ & $7.8\times10^{-5}$ & $4.4\times10^{-4}$ & $4.0\times10^{-3}$ & $5.0\times10^{-4}$ & $2.5\times10^{-3}$ & $3.1\times10^{-3}$ & $0$ & $0$ & $0$ \\
        $^{14}$N & $1.6\times10^{-2}$ & $1.8\times10^{-2}$ & $1.1\times10^{-2}$ & $3.6\times10^{-3}$ & $9.0\times10^{-3}$ & $1.1\times10^{-2}$ & $8.1\times10^{-3}$ & $0$ & $0$ & $0$ \\
        $^{15}$N & $4.1\times10^{-5}$ & $6.2\times10^{-4}$ & $7.0\times10^{-4}$ & $1.8\times10^{-4}$ & $2.1\times10^{-3}$ & $1.1\times10^{-4}$ & $3.9\times10^{-4}$ & $0$ & $0$ & $0$ \\
        $^{16}$O & $1.1\times10^{-1}$ & $2.4\times10^{-1}$ & $1.50$ & $7.1\times10^{-1}$ & $2.15$ & $1.52$ & $1.10$ & $0$ & $0$ & $0$ \\
        $^{17}$O & $1.9\times10^{-5}$ & $1.7\times10^{-5}$ & $2.9\times10^{-5}$ & $1.1\times10^{-5}$ & $7.6\times10^{-4}$ & $5.5\times10^{-5}$ & $3.3\times10^{-5}$ & $0$ & $0$ & $0$ \\
        $^{18}$O & $3.0\times10^{-5}$ & $2.6\times10^{-5}$ & $4.6\times10^{-3}$ & $8.3\times10^{-6}$ & $4.2\times10^{-6}$ & $3.9\times10^{-7}$ & $4.7\times10^{-7}$ & $0$ & $0$ & $0$ \\
        $^{18}$F & $0$ & $0$ & $0$ & $1.1\times10^{-8}$ & $7.8\times10^{-6}$ & $3.0\times10^{-9}$ & $3.2\times10^{-8}$ & $0$ & $0$ & $0$ \\
        $^{19}$F & $1.3\times10^{-6}$ & $1.3\times10^{-6}$ & $5.2\times10^{-5}$ & $1.2\times10^{-7}$ & $3.4\times10^{-6}$ & $1.8\times10^{-6}$ & $1.1\times10^{-6}$ & $0$ & $0$ & $0$ \\
        $^{20}$Ne & $5.9\times10^{-5}$ & $1.1\times10^{-3}$ & $3.0\times10^{-2}$ & $1.6\times10^{-1}$ & $1.4\times10^{-1}$ & $2.2\times10^{-1}$ & $2.0\times10^{-2}$ & $0$ & $0$ & $0$ \\
        $^{21}$Ne & $7.4\times10^{-7}$ & $4.7\times10^{-6}$ & $3.0\times10^{-4}$ & $8.3\times10^{-7}$ & $2.5\times10^{-4}$ & $3.0\times10^{-5}$ & $1.4\times10^{-5}$ & $0$ & $0$ & $0$ \\
        $^{22}$Ne & $1.9\times10^{-4}$ & $1.9\times10^{-4}$ & $2.1\times10^{-2}$ & $2.3\times10^{-5}$ & $2.2\times10^{-4}$ & $3.4\times10^{-3}$ & $6.5\times10^{-4}$ & $0$ & $0$ & $0$ \\
        $^{23}$Na & $1.3\times10^{-6}$ & $1.5\times10^{-6}$ & $1.8\times10^{-4}$ & $5.1\times10^{-7}$ & $3.7\times10^{-6}$ & $3.5\times10^{-5}$ & $1.1\times10^{-5}$ & $0$ & $0$ & $0$ \\
        $^{24}$Mg & $1.5\times10^{-6}$ & $5.7\times10^{-5}$ & $3.0\times10^{-3}$ & $9.0\times10^{-3}$ & $1.5\times10^{-2}$ & $1.5\times10^{-2}$ & $8.5\times10^{-4}$ & $0$ & $0$ & $0$ \\
        $^{25}$Mg & $1.8\times10^{-6}$ & $1.4\times10^{-5}$ & $2.2\times10^{-3}$ & $6.9\times10^{-7}$ & $1.2\times10^{-3}$ & $6.3\times10^{-5}$ & $1.3\times10^{-5}$ & $0$ & $0$ & $0$ \\
        $^{26}$Mg & $4.8\times10^{-6}$ & $2.2\times10^{-5}$ & $5.4\times10^{-3}$ & $1.2\times10^{-6}$ & $4.6\times10^{-4}$ & $1.4\times10^{-4}$ & $1.6\times10^{-5}$ & $0$ & $0$ & $0$ \\
        $^{26}$Al & $0$ & $0$ & $0$ & $0$ & $0$ & $7.9\times10^{-10}$ & $0$ & $0$ & $0$ & $0$ \\
        $^{27}$Al & $3.0\times10^{-8}$ & $9.5\times10^{-8}$ & $3.9\times10^{-5}$ & $7.2\times10^{-9}$ & $1.7\times10^{-6}$ & $8.8\times10^{-7}$ & $6.8\times10^{-8}$ & $0$ & $0$ & $0$ \\
        $^{28}$Si & $0$ & $0$ & $0$ & $1.8\times10^{-4}$ & $7.9\times10^{-4}$ & $3.7\times10^{-4}$ & $1.1\times10^{-5}$ & $0$ & $0$ & $0$ \\
        $^{29}$Si & $0$ & $0$ & $0$ & $0$ & $0$ & $0$ & $0$ & $0$ & $0$ & $0$ \\
        $^{30}$Si & $0$ & $0$ & $0$ & $0$ & $0$ & $0$ & $0$ & $0$ & $0$ & $0$ \\
        $^{31}$P & $0$ & $0$ & $0$ & $0$ & $0$ & $0$ & $0$ & $0$ & $0$ & $0$ \\
        $^{32}$S & $0$ & $0$ & $0$ & $1.8\times10^{-7}$ & $2.4\times10^{-5}$ & $7.0\times10^{-7}$ & $1.6\times10^{-8}$ & $0$ & $0$ & $0$ \\
        $^{33}$S & $0$ & $0$ & $0$ & $0$ & $0$ & $0$ & $0$ & $0$ & $0$ & $0$ \\
        $^{34}$S & $0$ & $0$ & $0$ & $0$ & $0$ & $0$ & $0$ & $0$ & $0$ & $0$ \\
        $^{36}$S & $0$ & $0$ & $0$ & $0$ & $0$ & $0$ & $0$ & $0$ & $0$ & $0$ \\
        $^{35}$Cl & $0$ & $0$ & $0$ & $0$ & $0$ & $0$ & $0$ & $0$ & $0$ & $0$ \\
        $^{37}$Cl & $0$ & $0$ & $0$ & $0$ & $0$ & $0$ & $0$ & $0$ & $0$ & $0$ \\
        $^{36}$Ar & $0$ & $0$ & $0$ & $0$ & $7.1\times10^{-8}$ & $0$ & $0$ & $0$ & $0$ & $0$ \\
        $^{38}$Ar & $0$ & $0$ & $0$ & $0$ & $0$ & $0$ & $0$ & $0$ & $0$ & $0$ \\
        $^{39}$K & $0$ & $0$ & $0$ & $0$ & $0$ & $0$ & $0$ & $0$ & $0$ & $0$ \\
        $^{40}$K & $0$ & $0$ & $0$ & $0$ & $0$ & $0$ & $0$ & $0$ & $0$ & $0$ \\
        $^{41}$K & $0$ & $0$ & $0$ & $0$ & $0$ & $0$ & $0$ & $0$ & $0$ & $0$ \\
        $^{40}$Ca & $0$ & $0$ & $0$ & $0$ & $0$ & $0$ & $0$ & $0$ & $0$ & $0$ \\
        $^{42}$Ca & $0$ & $0$ & $0$ & $0$ & $0$ & $0$ & $0$ & $0$ & $0$ & $0$ \\
        $^{43}$Ca & $0$ & $0$ & $0$ & $0$ & $0$ & $0$ & $0$ & $0$ & $0$ & $0$ \\
        $^{44}$Ca & $0$ & $0$ & $0$ & $0$ & $0$ & $0$ & $0$ & $0$ & $0$ & $0$ \\
        $^{45}$Sc & $0$ & $0$ & $0$ & $0$ & $0$ & $0$ & $0$ & $0$ & $0$ & $0$ \\
        $^{46}$Ti & $0$ & $0$ & $0$ & $0$ & $0$ & $0$ & $0$ & $0$ & $0$ & $0$ \\
        $^{47}$Ti & $0$ & $0$ & $0$ & $0$ & $0$ & $0$ & $0$ & $0$ & $0$ & $0$ \\
        $^{48}$Ti & $0$ & $0$ & $0$ & $0$ & $0$ & $0$ & $0$ & $0$ & $0$ & $0$ \\
        $^{49}$Ti & $0$ & $0$ & $0$ & $0$ & $0$ & $0$ & $0$ & $0$ & $0$ & $0$ \\
        $^{51}$V & $0$ & $0$ & $0$ & $0$ & $0$ & $0$ & $0$ & $0$ & $0$ & $0$ \\
        $^{50}$Cr & $0$ & $0$ & $0$ & $0$ & $0$ & $0$ & $0$ & $0$ & $0$ & $0$ \\
        $^{52}$Cr & $0$ & $0$ & $0$ & $0$ & $0$ & $0$ & $0$ & $0$ & $0$ & $0$ \\
        $^{53}$Cr & $0$ & $0$ & $0$ & $0$ & $0$ & $0$ & $0$ & $0$ & $0$ & $0$ \\
        $^{54}$Cr & $0$ & $0$ & $0$ & $0$ & $0$ & $0$ & $0$ & $0$ & $0$ & $0$ \\
        $^{55}$Mn & $0$ & $0$ & $0$ & $0$ & $0$ & $0$ & $0$ & $0$ & $0$ & $0$ \\
        $^{54}$Fe & $0$ & $0$ & $0$ & $0$ & $0$ & $0$ & $0$ & $0$ & $0$ & $0$ \\
        $^{56}$Fe & $0$ & $0$ & $0$ & $0$ & $0$ & $0$ & $0$ & $0$ & $0$ & $0$ \\
        $^{57}$Fe & $0$ & $0$ & $0$ & $0$ & $0$ & $0$ & $0$ & $0$ & $0$ & $0$ \\
        $^{58}$Fe & $0$ & $0$ & $0$ & $0$ & $0$ & $0$ & $0$ & $0$ & $0$ & $0$ \\
        $^{59}$Co & $0$ & $0$ & $0$ & $0$ & $0$ & $0$ & $0$ & $0$ & $0$ & $0$ \\
        $^{58}$Ni & $0$ & $0$ & $0$ & $0$ & $0$ & $0$ & $0$ & $0$ & $0$ & $0$ \\
        $^{60}$Ni & $0$ & $0$ & $0$ & $0$ & $0$ & $0$ & $0$ & $0$ & $0$ & $0$ \\
        $^{61}$Ni & $0$ & $0$ & $0$ & $0$ & $0$ & $0$ & $0$ & $0$ & $0$ & $0$ \\
        $^{62}$Ni & $0$ & $0$ & $0$ & $0$ & $0$ & $0$ & $0$ & $0$ & $0$ & $0$ \\
        $^{63}$Cu & $0$ & $0$ & $0$ & $0$ & $0$ & $0$ & $0$ & $0$ & $0$ & $0$ \\
        $^{64}$Zn & $0$ & $0$ & $0$ & $0$ & $0$ & $0$ & $0$ & $0$ & $0$ & $0$ \\
        $^{66}$Zn & $0$ & $0$ & $0$ & $0$ & $0$ & $0$ & $0$ & $0$ & $0$ & $0$ \\
        \hline
    \end{tabular}
    }
\end{table*}

\begin{table*}[htbp]
    \caption{Gross adopted ejecta masses for selected isotopes} for individual isotopes as a function of initial stellar mass ($M_\odot$). For $60 \leq M/M_\odot \leq 90$, only pre-SN stellar wind ejecta are included (terminal SN ejecta for this range are assumed to fall into the remnant); outside this range, total ejecta = terminal SN/PPISN/PISN + pre-SN stellar winds. Entries marked `$0$' fall below $10^{-10}\,M_\odot$.
    \label{tab:isotope_yields_windonly:p2}
    \scriptsize                 
    \setlength{\tabcolsep}{6pt}    
    \renewcommand{\arraystretch}{1}      
    \resizebox{\textwidth}{!}{
    \hspace*{-0.17\textwidth}
    \begin{tabular}{|l|cccccccccc|}
        \hline
        Isotope & \multicolumn{10}{c|}{Stellar Mass ($M_\odot$)} \\
        \cline{2-11} & $60$ & $65$ & $70$ & $80$ & $90$ & $100$ & $120$ & $150$ & $180$ & $200$ \\
        \hline
        $^{1}$H & $1.05$ & $1.67$ & $1.57$ & $1.86$ & $2.13$ & $18.2$ & $19.3$ & $20.8$ & $21.5$ & $22.9$ \\
        $^{3}$He & $0$ & $0$ & $0$ & $0$ & $0$ & $0$ & $0$ & $0$ & $0$ & $0$ \\
        $^{4}$He & $3.8\times10^{-1}$ & $6.0\times10^{-1}$ & $5.6\times10^{-1}$ & $7.2\times10^{-1}$ & $8.0\times10^{-1}$ & $28.7$ & $38.0$ & $49.6$ & $57.6$ & $55.8$ \\
        $^{12}$C & $0$ & $0$ & $0$ & $0$ & $0$ & $8.6\times10^{-1}$ & $4.21$ & $11.3$ & $9.00$ & $6.89$ \\
        $^{13}$C & $0$ & $0$ & $0$ & $0$ & $0$ & $1.7\times10^{-3}$ & $4.8\times10^{-1}$ & $6.5\times10^{-1}$ & $8.0\times10^{-1}$ & $6.7\times10^{-1}$ \\
        $^{14}$C & $0$ & $0$ & $0$ & $0$ & $0$ & $0$ & $0$ & $0$ & $0$ & $0$ \\
        $^{14}$N & $0$ & $0$ & $0$ & $0$ & $0$ & $8.9\times10^{-3}$ & $1.23$ & $1.81$ & $2.52$ & $2.42$ \\
        $^{15}$N & $0$ & $0$ & $0$ & $0$ & $0$ & $0$ & $0$ & $7.3\times10^{-7}$ & $6.6\times10^{-6}$ & $6.4\times10^{-6}$ \\
        $^{16}$O & $0$ & $0$ & $0$ & $0$ & $0$ & $5.89$ & $9.50$ & $57.0$ & $55.6$ & $65.4$ \\
        $^{17}$O & $0$ & $0$ & $0$ & $0$ & $0$ & $1.4\times10^{-5}$ & $1.4\times10^{-3}$ & $2.1\times10^{-3}$ & $3.0\times10^{-3}$ & $1.1\times10^{-2}$ \\
        $^{18}$O & $0$ & $0$ & $0$ & $0$ & $0$ & $2.6\times10^{-5}$ & $1.7\times10^{-6}$ & $9.8\times10^{-6}$ & $1.4\times10^{-5}$ & $1.7\times10^{-5}$ \\
        $^{18}$F & $0$ & $0$ & $0$ & $0$ & $0$ & $0$ & $0$ & $0$ & $0$ & $0$ \\
        $^{19}$F & $0$ & $0$ & $0$ & $0$ & $0$ & $4.3\times10^{-9}$ & $5.7\times10^{-8}$ & $1.0\times10^{-4}$ & $1.1\times10^{-5}$ & $5.7\times10^{-7}$ \\
        $^{20}$Ne & $0$ & $0$ & $0$ & $0$ & $0$ & $5.8\times10^{-1}$ & $1.4\times10^{-1}$ & $5.01$ & $4.22$ & $4.26$ \\
        $^{21}$Ne & $0$ & $0$ & $0$ & $0$ & $0$ & $3.7\times10^{-5}$ & $2.4\times10^{-5}$ & $7.4\times10^{-4}$ & $5.7\times10^{-4}$ & $5.5\times10^{-4}$ \\
        $^{22}$Ne & $0$ & $0$ & $0$ & $0$ & $0$ & $4.0\times10^{-5}$ & $1.3\times10^{-4}$ & $1.1\times10^{-3}$ & $1.5\times10^{-3}$ & $7.4\times10^{-4}$ \\
        $^{23}$Na & $0$ & $0$ & $0$ & $0$ & $0$ & $1.8\times10^{-6}$ & $3.6\times10^{-6}$ & $9.1\times10^{-3}$ & $3.0\times10^{-3}$ & $2.8\times10^{-3}$ \\
        $^{24}$Mg & $0$ & $0$ & $0$ & $0$ & $0$ & $1.4\times10^{-1}$ & $1.4\times10^{-2}$ & $1.53$ & $3.79$ & $4.40$ \\
        $^{25}$Mg & $0$ & $0$ & $0$ & $0$ & $0$ & $1.5\times10^{-4}$ & $4.4\times10^{-5}$ & $3.8\times10^{-3}$ & $2.7\times10^{-3}$ & $5.5\times10^{-3}$ \\
        $^{26}$Mg & $0$ & $0$ & $0$ & $0$ & $0$ & $9.0\times10^{-5}$ & $4.8\times10^{-5}$ & $3.8\times10^{-3}$ & $1.9\times10^{-3}$ & $2.9\times10^{-3}$ \\
        $^{26}$Al & $0$ & $0$ & $0$ & $0$ & $0$ & $7.1\times10^{-9}$ & $8.3\times10^{-9}$ & $0$ & $0$ & $0$ \\
        $^{27}$Al & $0$ & $0$ & $0$ & $0$ & $0$ & $8.9\times10^{-8}$ & $1.0\times10^{-7}$ & $3.4\times10^{-2}$ & $1.6\times10^{-2}$ & $1.7\times10^{-2}$ \\
        $^{28}$Si & $0$ & $0$ & $0$ & $0$ & $0$ & $5.9\times10^{-10}$ & $1.4\times10^{-4}$ & $3.2\times10^{-1}$ & $14.8$ & $21.7$ \\
        $^{29}$Si & $0$ & $0$ & $0$ & $0$ & $0$ & $0$ & $0$ & $2.0\times10^{-3}$ & $2.2\times10^{-2}$ & $2.0\times10^{-2}$ \\
        $^{30}$Si & $0$ & $0$ & $0$ & $0$ & $0$ & $0$ & $0$ & $6.5\times10^{-3}$ & $1.3\times10^{-3}$ & $1.3\times10^{-3}$ \\
        $^{31}$P & $0$ & $0$ & $0$ & $0$ & $0$ & $0$ & $0$ & $2.7\times10^{-3}$ & $1.8\times10^{-3}$ & $1.3\times10^{-3}$ \\
        $^{32}$S & $0$ & $0$ & $0$ & $0$ & $0$ & $0$ & $8.1\times10^{-7}$ & $9.5\times10^{-4}$ & $5.27$ & $9.06$ \\
        $^{33}$S & $0$ & $0$ & $0$ & $0$ & $0$ & $0$ & $0$ & $3.8\times10^{-5}$ & $3.7\times10^{-3}$ & $3.7\times10^{-3}$ \\
        $^{34}$S & $0$ & $0$ & $0$ & $0$ & $0$ & $0$ & $0$ & $2.1\times10^{-4}$ & $2.5\times10^{-3}$ & $7.3\times10^{-4}$ \\
        $^{36}$S & $0$ & $0$ & $0$ & $0$ & $0$ & $0$ & $0$ & $1.1\times10^{-8}$ & $6.2\times10^{-9}$ & $3.7\times10^{-9}$ \\
        $^{35}$Cl & $0$ & $0$ & $0$ & $0$ & $0$ & $0$ & $0$ & $5.6\times10^{-5}$ & $7.0\times10^{-4}$ & $3.8\times10^{-4}$ \\
        $^{37}$Cl & $0$ & $0$ & $0$ & $0$ & $0$ & $0$ & $0$ & $1.1\times10^{-7}$ & $6.8\times10^{-4}$ & $6.8\times10^{-4}$ \\
        $^{36}$Ar & $0$ & $0$ & $0$ & $0$ & $0$ & $0$ & $3.8\times10^{-10}$ & $1.3\times10^{-5}$ & $6.8\times10^{-1}$ & $1.36$ \\
        $^{38}$Ar & $0$ & $0$ & $0$ & $0$ & $0$ & $0$ & $0$ & $4.9\times10^{-7}$ & $2.9\times10^{-3}$ & $8.3\times10^{-4}$ \\
        $^{39}$K & $0$ & $0$ & $0$ & $0$ & $0$ & $0$ & $0$ & $1.3\times10^{-7}$ & $9.0\times10^{-4}$ & $4.8\times10^{-4}$ \\
        $^{40}$K & $0$ & $0$ & $0$ & $0$ & $0$ & $0$ & $0$ & $2.5\times10^{-10}$ & $2.0\times10^{-8}$ & $9.2\times10^{-9}$ \\
        $^{41}$K & $0$ & $0$ & $0$ & $0$ & $0$ & $0$ & $0$ & $2.4\times10^{-10}$ & $1.5\times10^{-4}$ & $1.5\times10^{-4}$ \\
        $^{40}$Ca & $0$ & $0$ & $0$ & $0$ & $0$ & $0$ & $0$ & $1.5\times10^{-8}$ & $4.4\times10^{-1}$ & $1.04$ \\
        $^{42}$Ca & $0$ & $0$ & $0$ & $0$ & $0$ & $0$ & $0$ & $2.4\times10^{-10}$ & $7.5\times10^{-5}$ & $2.2\times10^{-5}$ \\
        $^{43}$Ca & $0$ & $0$ & $0$ & $0$ & $0$ & $0$ & $0$ & $0$ & $2.4\times10^{-8}$ & $9.5\times10^{-9}$ \\
        $^{44}$Ca & $0$ & $0$ & $0$ & $0$ & $0$ & $0$ & $0$ & $0$ & $9.6\times10^{-5}$ & $3.0\times10^{-4}$ \\
        $^{45}$Sc & $0$ & $0$ & $0$ & $0$ & $0$ & $0$ & $0$ & $0$ & $3.1\times10^{-6}$ & $3.5\times10^{-6}$ \\
        $^{46}$Ti & $0$ & $0$ & $0$ & $0$ & $0$ & $0$ & $0$ & $0$ & $4.1\times10^{-5}$ & $1.8\times10^{-5}$ \\
        $^{47}$Ti & $0$ & $0$ & $0$ & $0$ & $0$ & $0$ & $0$ & $0$ & $3.6\times10^{-7}$ & $6.8\times10^{-7}$ \\
        $^{48}$Ti & $0$ & $0$ & $0$ & $0$ & $0$ & $0$ & $0$ & $0$ & $9.4\times10^{-4}$ & $6.6\times10^{-3}$ \\
        $^{49}$Ti & $0$ & $0$ & $0$ & $0$ & $0$ & $0$ & $0$ & $0$ & $4.6\times10^{-5}$ & $1.9\times10^{-4}$ \\
        $^{51}$V & $0$ & $0$ & $0$ & $0$ & $0$ & $0$ & $0$ & $0$ & $8.4\times10^{-5}$ & $3.2\times10^{-4}$ \\
        $^{50}$Cr & $0$ & $0$ & $0$ & $0$ & $0$ & $0$ & $0$ & $0$ & $3.9\times10^{-4}$ & $5.0\times10^{-4}$ \\
        $^{52}$Cr & $0$ & $0$ & $0$ & $0$ & $0$ & $0$ & $0$ & $0$ & $1.4\times10^{-2}$ & $1.4\times10^{-1}$ \\
        $^{53}$Cr & $0$ & $0$ & $0$ & $0$ & $0$ & $0$ & $0$ & $0$ & $9.2\times10^{-4}$ & $5.7\times10^{-3}$ \\
        $^{54}$Cr & $0$ & $0$ & $0$ & $0$ & $0$ & $0$ & $0$ & $0$ & $9.7\times10^{-9}$ & $4.3\times10^{-9}$ \\
        $^{55}$Mn & $0$ & $0$ & $0$ & $0$ & $0$ & $0$ & $0$ & $0$ & $3.9\times10^{-3}$ & $2.3\times10^{-2}$ \\
        $^{54}$Fe & $0$ & $0$ & $0$ & $0$ & $0$ & $0$ & $0$ & $0$ & $4.4\times10^{-2}$ & $9.6\times10^{-2}$ \\
        $^{56}$Fe & $0$ & $0$ & $0$ & $0$ & $0$ & $0$ & $0$ & $0$ & $2.4\times10^{-1}$ & $3.51$ \\
        $^{57}$Fe & $0$ & $0$ & $0$ & $0$ & $0$ & $0$ & $0$ & $0$ & $1.8\times10^{-3}$ & $1.6\times10^{-2}$ \\
        $^{58}$Fe & $0$ & $0$ & $0$ & $0$ & $0$ & $0$ & $0$ & $0$ & $8.8\times10^{-9}$ & $6.4\times10^{-9}$ \\
        $^{59}$Co & $0$ & $0$ & $0$ & $0$ & $0$ & $0$ & $0$ & $0$ & $4.4\times10^{-6}$ & $7.0\times10^{-6}$ \\
        $^{58}$Ni & $0$ & $0$ & $0$ & $0$ & $0$ & $0$ & $0$ & $0$ & $2.8\times10^{-3}$ & $1.0\times10^{-2}$ \\
        $^{60}$Ni & $0$ & $0$ & $0$ & $0$ & $0$ & $0$ & $0$ & $0$ & $6.5\times10^{-6}$ & $1.1\times10^{-5}$ \\
        $^{61}$Ni & $0$ & $0$ & $0$ & $0$ & $0$ & $0$ & $0$ & $0$ & $9.2\times10^{-10}$ & $3.6\times10^{-7}$ \\
        $^{62}$Ni & $0$ & $0$ & $0$ & $0$ & $0$ & $0$ & $0$ & $0$ & $2.2\times10^{-10}$ & $1.5\times10^{-6}$ \\
        $^{63}$Cu & $0$ & $0$ & $0$ & $0$ & $0$ & $0$ & $0$ & $0$ & $0$ & $5.5\times10^{-10}$ \\
        $^{64}$Zn & $0$ & $0$ & $0$ & $0$ & $0$ & $0$ & $0$ & $0$ & $0$ & $8.1\times10^{-10}$ \\
        $^{66}$Zn & $0$ & $0$ & $0$ & $0$ & $0$ & $0$ & $0$ & $0$ & $0$ & $1.0\times10^{-9}$ \\
        \hline
    \end{tabular}
    }
\end{table*}


\end{document}